\documentclass[11pt,a4paper]{article}
\usepackage{jcappub}
\pdfoutput=1
\usepackage[english]{babel}
\usepackage{amsmath,amssymb,amsfonts, bm,bbm,slashed, subdepth}
\usepackage{graphicx}
\usepackage[sort&compress]{natbib}
\usepackage{xcolor}
\usepackage[normalem]{ulem}
\usepackage{hyperref}
\usepackage{cleveref}
\definecolor{red}{rgb}{1.0, 0, 0}
\usepackage{hyperref}
\usepackage{enumerate}
\usepackage{epsfig, subfigure}
\usepackage{setspace}
\usepackage{booktabs, tabularx}
\usepackage{units}
\usepackage{hyperref}
\usepackage{subfigure}
\mathchardef\mhyphen="2D 

\newcommand{\be}{\begin{equation}}
\newcommand{\ee}{\end{equation}}
\newcommand{\ba}{\begin{array}}
\newcommand{\ea}{\end{array}}
\newcommand{\bea}{\begin{eqnarray}}
\newcommand{\eea}{\end{eqnarray}}
\newcommand{\balg}{\begin{align}}
\newcommand{\ealg}{\end{align}}
\newcommand{\bit}{\begin{itemize}}
\newcommand{\eit}{\end{itemize}}
\newcommand{\trm}[1]{\textrm{#1}}

\newcommand{\Mpc}{\trm{\Mpc}}
\newcommand{\yr}{\trm{\yr}}
\newcommand{\eV}{\trm{\eV}}

\usepackage{pifont}

\begin{document}

\title{The Effect of Thermal Velocities on Structure Formation in N-body Simulations of Warm Dark Matter}
\author[a]{Matteo Leo,}
\author[b]{Carlton M. Baugh,}
\author[b]{Baojiu Li}
\author[a]{and Silvia Pascoli}
\affiliation[a]{Institute for Particle Physics Phenomenology, Department of Physics, Durham University, South Road, Durham DH1 3LE, U.K.}
\affiliation[b]{Institute for Computational Cosmology, Department of Physics, Durham University, South Road, Durham DH1 3LE, U.K.}
\emailAdd{matteo.leo@durham.ac.uk, c.m.baugh@durham.ac.uk, baojiu.li@durham.ac.uk, silvia.pascoli@durham.ac.uk}
\hfill{IPPP/17/52}

\abstract{We investigate the impact of thermal velocities in N-body simulations of structure formation in warm dark matter models. Adopting the commonly used approach of adding thermal velocities, randomly selected from a Fermi-Dirac distribution, to the gravitationally-induced velocities of the simulation particles, we compare the matter and velocity power spectra measured from CDM and WDM simulations, in the latter case with and without thermal velocities. This prescription for adding thermal velocities introduces numerical noise into the initial conditions, which influences structure formation. At early times, the noise affects dramatically the power spectra measured from simulations with thermal velocities, with deviations of the order of $\sim \mathcal{O}(10)$ (in the matter power spectra) and of the order of $\sim \mathcal{O}(10^2)$ (in the velocity power spectra) compared to those extracted from simulations without thermal velocities. At late times, these effects are less pronounced with deviations of less than a few percent. Increasing the resolution of the N-body simulation shifts these discrepancies to higher wavenumbers. We also find that spurious haloes start to appear in simulations which include thermal velocities at a mass that is $\sim$3 times larger than in simulations without thermal velocities.}

\maketitle

\section{Introduction}
Although the cold dark matter (CDM) paradigm successfully reproduces the observed structure of the Universe on large and intermediate scales, several studies have suggested tensions between pure CDM predictions and observations on small scales (for a recent review of the small scale problems in CDM see \cite{Weinberg:2013aya}). CDM haloes display central density cusps in N-body simulations \cite{Dubinski:1991bm,Navarro:1995iw,Navarro:1996gj}. Such profiles are strongly excluded by the observed small scale dynamics of some spiral galaxies, which in turn seem to prefer a constant DM distribution (core) in the center \cite{deBlok:2001hbg, Salucci:2011ee}. The ``missing satellites'' problem is another small scale issue. This refers to the large difference between the number of satellite galaxies observed in Milky Way-like galaxies and the number of subhaloes in CDM simulations \cite{Klypin:1999uc,Moore:1999nt}. It is unclear if this problem is really due to an absence of small structures or rather that these structures are ``dark'' haloes which contain no visible galaxies and hence are not detectable directly.  Several solutions have been proposed to ameliorate these shortcomings within the CDM paradigm, e.g. by taking into account baryonic physics \cite{Mashchenko:2007jp,2012MNRAS.421.3464P,2013MNRAS.432.1947M,2014ApJ...786...87B}. The absence of massive subhaloes in the Milky Way could be also interpreted as an indication that the MW halo is less massive than is commonly assumed \cite{2012MNRAS.424.2715W}. In any case, the observed lack of small structures implies that galaxy formation takes place in the most massive MW subhaloes, but when we look at these structures they appear less dense than expected in CDM simulations. This is the so-called ``too big to fail'' problem, first identified by \cite{2011MNRAS.415L..40B}, although this issue can also be resolved by invoking  baryonic physics \cite{2014ApJ...786...87B}. 

The difficulties facing CDM have renewed interest in alternative DM scenarios that display less power on small scales. These scenarios are commonly grouped under the name of ``non-cold'' dark matter (hereafter nCDM). The physical mechanisms leading to the suppression of the gravitational clustering on small scales depend on the particular particle production process, but usually nCDM candidates are characterised by either non-negligible thermodynamic velocity dispersion \cite{Bode:2000gq,Colin:2000dn,Hansen:2001zv,Viel:2005qj,Dodelson:1993je,Dolgov:2000ew,Asaka:2006nq,Enqvist:1990ek,Shi:1998km,Abazajian:2001nj,Kusenko:2006rh,Petraki:2007gq,Merle:2015oja,Konig:2016dzg}, interactions (DM interacting with standard model particles such as neutrinos or photons \cite{Boehm:2004th,Boehm:2014vja} and self-interacting DM \cite{Spergel:1999mh}) or pressure terms from macroscopic wave-like behaviour (e.g. ultra-light axions \cite{Marsh:2015xka}). These features introduce a characteristic scale below which the density fluctuations are erased. For example, the main consequence of the presence of a non-vanishing velocity dispersion is free-streaming: the particles cannot be confined into regions smaller than the free-streaming length, $\lambda_{fs} \sim \sigma/H$, where $\sigma$ is the particle velocity dispersion, so that density perturbations are damped on scales below  $\lambda_{fs}$ and a cut-off appears in the linear matter power spectrum. The shape of the matter power spectrum and the position of the cut-off depends on the velocity dispersion $\sigma$, which in turn depends on the particular velocity distribution of the particles. The free-streaming is a feature which is common in collisionless fluids with non-negligible velocity dispersions, e.g. active neutrinos  (see e.g.  \cite{Lesgourgues:2006nd}) exhibit such behaviour\footnote{CDM candidates such as WIMPs also display a cut-off in the power spectrum due to free-streaming or to interactions in the early Universe. Since such particles  are heavy and interact weakly with Standard Model particles, the suppression scale appears at very small masses (between $10^{-12}$ $h^{-1}$M$_\odot$ and $10^{-4}$ $h^{-1}$M$_\odot$ depending on the parameters of the model \cite{Green:2005kf,Profumo:2006bv}). }. 

Several nCDM models have been proposed in which candidates have non-negligible thermodynamical velocities. The simplest scenario is the so-called thermal warm dark matter (WDM), where DM particles are in thermal contact with the primordial bath, decouple when still relativistic, and enter the non-relativistic regime deep into the radiation-dominated period. This mechanism ensures that the particles have a Fermi-Dirac (if the particles are fermions) momentum distribution, which leads to a simple relation between the particle mass and the free-streaming length $\lambda_{fs}$, and a distinctive cut-off in the matter power spectrum \cite{Bode:2000gq,Colin:2000dn,Hansen:2001zv,Viel:2005qj}. The Lyman-$\alpha$ forest has proven to be a powerful tool to constrain the mass of the thermal WDM candidates, suggesting a mass of $m_\mathrm{WDM} \gtrsim 3.3$ keV \cite{Viel:2013apy} (although a recent analysis pushes the limit to slightly higher masses $m_\mathrm{WDM} \gtrsim 3.5$ keV \cite{Irsic:2017ixq}). These constraints on the mass are strictly valid only if all the DM sector is in the form of thermal WDM particles. Mixed scenarios (thermal WDM + CDM) are also possible. In these models only a fraction of the dark matter is in the form of WDM, while the remaining fraction behaves as standard CDM. Such models do not display the sharp cut-off in the power spectrum that is characteristic of a pure WDM model since the CDM fluctuations are present on small scales \cite{Boyarsky:2008xj}. Constraints on mixed DM using Lyman-$\alpha$ forest have been found in \cite{Boyarsky:2008xj}.

keV sterile neutrinos are another class of nCDM candidates, which are well motivated by particle physics theories (see \cite{Adhikari:2016bei} for a summary of the cosmological impact of these particles). Their mixing with active neutrinos offers a possible mechanism for producing a primordial abundance of such particles. Moreover, sterile neutrinos $\nu_s$ produced via mixing can  decay radiatively via $\nu_s \rightarrow \nu_a +\gamma$, where $\nu_a$ is an active neutrino, leading to a monochromatic X-ray line. The mass and the mixing angle of these particles can then be constrained by searching for decay signals in the X-ray spectrum \cite{Adhikari:2016bei,Dolgov:2000ew}.   The production mechanism via mixing can be non-resonant (``Dodelson-Widrow'') \cite{Dodelson:1993je,Dolgov:2000ew, Asaka:2006nq} or resonantly (``Shi-Fuller'') enhanced by a lepton asymmetry in the early Universe \cite{Enqvist:1990ek,Shi:1998km,Abazajian:2001nj}. However, a pure non-resonant production mechanism of keV sterile neutrinos is in conflict with the current astrophysical constraints from the Lyman-$\alpha$ and X-rays \cite{Adhikari:2016bei,Viel:2013apy}. Resonantly produced sterile neutrinos are more in line with observations and can potentially explain the recent (although controversial) detection of a $3.55$ keV X-ray emission line \cite{Bulbul:2014sua,Boyarsky:2014jta}. However, such neutrinos are constrained to a very narrow range of masses $7 \leq m_{s} \leq 36$ keV and lepton asymmetries larger than $15 \times 10^{-6}$ (at $95\%$ CL) \cite{Cherry:2017dwu}, and they are completely disfavoured when including constraints from Lyman-$\alpha$ forest \cite{Schneider:2016uqi}. Another possibility is that keV sterile neutrinos can be produced via the decay of a hypothetical parent particle in the early Universe \cite{Kusenko:2006rh,Petraki:2007gq,Merle:2015oja,Konig:2016dzg}. The initial parent decay production can be followed by a non-resonant production \cite{Merle:2015vzu}. This case is similar to a mixed DM scenario since there is a colder velocity component (spectrum from decay production) and a warmer velocity component (Dodelson-Widrow production). Sterile neutrinos produced by mixing or by particle decays have non-thermal distributions and cannot be treated in the simple framework of thermal WDM particles. Indeed, in some of these models, the sterile neutrino momentum distribution deviates significantly  from Fermi-Dirac, so that, unlike thermal relics, the matter power spectrum cut-off is not a simple function of the mass. Consequently, keV sterile neutrinos can act as warm or cold DM depending on the model thus eluding the $m_\mathrm{WDM} \gtrsim 3.3$ keV Lyman-$\alpha$ constraint found in  \cite{Viel:2013apy,Irsic:2017ixq} (which consider only matter power spectra from thermal WDM distributions). The very specific shape of the matter power spectrum has to be taken into account to constrain each (non-thermal) model using the Lyman-$\alpha$ forest. Another way to constrain non-thermal models using the Lyman-$\alpha$ forest has been proposed in \cite{Murgia:2017lwo}.

If DM is in the form of nCDM candidates, it is important to investigate how the predictions for structure formation differ in such models from those forecast in standard CDM. N-body simulations have proved to be a powerful tool to follow the evolution of the structure in the CDM scenario and can also be used to study the effects of the damping on small scales in nCDM. However, the difference between nCDM candidates, such as thermal relics or sterile neutrinos, and CDM is not only in the shape of the linear matter power spectrum, but also in the presence of non-negligible thermodynamical velocities for nCDM. Indeed, when running a simulation of structure formation, the starting redshift is generally chosen to be at a reasonably high value, to ensure that all of the scales we are interested in are well within the linear regime. If we start at $z=199$, for a typical simulation the gravitationally-induced velocities (generally called peculiar velocities) are around or below $v_\mathrm{pec} \sim 10$ km/s. CDM candidates have negligible thermodynamical velocities respect to the peculiar velocities $v_\mathrm{pec}$ at the redshift considered, and so their thermal velocities can be safely neglected. This is not always the case for nCDM candidates, e.g. for a thermal WDM candidate with mass of $3.3$ keV, the thermal velocity dispersion is of the order $\sigma_\mathrm{therm} \sim 2$ km/s at  $z = 199$ \cite{Bode:2000gq}. So, thermal velocities are non-negligible and should be taken into account in the simulations. However, numerical simulations of nCDM are usually initialised by  taking into account only the cut-off in the linear matter power spectrum, while thermal velocities are not explicitly included (see e.g. \cite{Bode:2000gq,Murgia:2017lwo,2012MNRAS.424..684S,2012MNRAS.420.2318L,Schneider:2013ria,Lovell:2013ola,Power:2013rpw,Power:2016usj,Bose:2015mga}). This is because thermal velocities are assumed to have no influence on the halo mass function at late times so long as the mass resolution of the N-body simulation at the starting redshift is well above the Jeans mass of the warm particle fluid \cite{Schneider:2013ria} (if a simulation belongs to this class, we say that the Jeans mass criterion is satisfied). Another argument that is often invoked to justify neglecting thermal velocities in N-body simulations concerns the distance travelled by particles free-streaming due to their thermal motions over the duration of the calculation, which is e.g. around a few kiloparsecs for thermal WDM candidates with masses around a few keV \cite{2012MNRAS.420.2318L}. If this length is smaller than the mean-interparticle separation in the simulation,  thermal velocities are neglected since free-streaming only influences scales that are not well resolved in the simulation (if a simulation belongs to this class, we say that the free-streaming length criterion is satisfied). Anyway, neglecting thermal velocities remains an approximation which we test here. The only physical effect expected when introducing primordial thermal velocities is a ``phase-packing'' limit, which prevents the density in the central regions of the haloes from increasing arbitrarily (producing a central core) \cite{Hogan:2000bv,2013MNRAS.428..882M,2013MNRAS.430.2346S}, but for values of WDM candidate masses compatible with the upper limits from the Ly-$\alpha$ forest \cite{Viel:2013apy, Irsic:2017ixq}, the cores are only a few parsecs in size and therefore are not relevant astrophysically \cite{2013MNRAS.428..882M,2013MNRAS.430.2346S}.  However, there are other models in which thermal velocities cannot be easily discarded, e.g. in the case of composite DM simulations, in which there are two DM components (e.g. the simulations of CDM + massive $\nu$ discussed in \cite{Brandbyge:2008rv, Viel:2010bn,2013JCAP...03..019V}). 
Our aim here is to assess the impact of including thermal velocities in simulations of structure formation in thermal WDM, using the standard technique described below, to see if this affects the range of scales on which the simulation results are accurate. 

A common approach used to implement thermal velocities in N-body simulations consists of adding the physical thermal velocities to the gravitationally induced peculiar velocities of the simulation particles in the N-body initial conditions (ICs) \cite{Boyarsky:2008xj,Colin:2007bk,Klypin:1992sf,Brandbyge:2008rv,Viel:2010bn,2013JCAP...03..019V,2012MNRAS.421...50V,2013MNRAS.428..882M,2013MNRAS.430.2346S,Paduroiu:2015jfa}. The subsequent evolution of structure will then follow both sources of the velocity field\footnote{A different approach to simulate gravitational evolution of collisionless WDM particles has been recently proposed in \cite{Hahn:2015sia}.}. This approach is also used in galactic dynamics, where the fluid is made up of stars. In this case it is well known that ignoring the stellar velocity dispersion produces gravitational instability (see e.g. \cite{Toomre:1964zx}). Simulations of galactic dynamics are performed by adding a stellar velocity, drawn from a distribution, to the simulation particles. However, as pointed out by \cite{Colin:2007bk,Klypin:1992sf,Viel:2010bn,Brandbyge:2008rv} (see also the discussion in \cite{Power:2013rpw,Power:2016usj}), the above approach of adding thermal velocities in the ICs can introduce a new source of numerical error. Indeed, since a simulation particle contains a huge number of physical particles\footnote{Simulation particles are always many orders of magnitude more massive than physical particles, e.g. in our simulations the simulation particle mass is around $10^7$ $h^{-1}$M$_\odot$, while the warm particle mass is of the order of keV $\sim 10^{-63}$ $h^{-1}$M$_\odot$. So each simulation particle is made up of around $10^{70}$ physical particles.}, the net thermal velocity has to be zero if the thermal velocities of physical particles are drawn from a distribution without a preferred direction. \cite{Colin:2007bk} pointed out that the numerical noise can be reduced by choosing a lower ($z\sim 40$) redshift at which to generate the ICs. However, a detailed study of how the noise affects the formation  and the evolution of the structures has not been carried out before.

Our aim is to perform a comprehensive study of the effects of thermal velocities in N-body simulations for the simplest scenario of thermal WDM where the velocity distribution is Fermi-Dirac\footnote{We stress here the fact that our simulations of thermal WDM satisfy both the Jeans mass \cite{Schneider:2013ria} and the free-streaming length criteria \cite{2012MNRAS.420.2318L}, so, in principle, we can also apply the approximation of neglecting thermal velocities. However, since we are interested in quantifying the effects of the noise introduced in the simulations when adding thermal velocities, we will take into account thermal velocities with the justification that at the initial high redshift, $z_\mathrm{ini}=199$, at which the ICs are generated, thermal velocities are comparable with peculiar ones.
}.  Using the approximation of assigning thermal velocities to the simulation particles, we quantify the impact of the numerical noise by analysing the matter and velocity power spectra measured from simulations at different redshifts. We also show how these numerical artefacts can affect halo properties such as the halo mass function and radial density profiles. In particular, we derive a new mass cut-off below which spurious haloes dominate in simulations with thermal velocities. Note also that the non-thermal production of sterile neutrinos produces, in general, colder velocity spectra than thermal WDM. For this reason, our results can be considered as an upper bound on the impact of thermal velocities in sterile neutrino models. 

The paper is organised as follows. In Section 2, we measure velocity power spectra for ICs generated at $z=199$ for WDM simulations, quantifying the source of the noise introduced by the thermal velocities in the initial conditions. We run a series of WDM simulations with thermal velocities varying the number of particles, box size and the physical WDM particle mass, in all cases extracting velocity power spectra and comparing with WDM simulations without thermal velocities.  Section 3 is devoted to the evolution of the ICs at $z=199$ up to $z=0$, showing how the noise in the ICs propagates through to subsequent times. We also discuss the possibility of reducing the numerical noise by generating the ICs at a lower redshift, $z=39$. In Section 4, we explore how haloes are affected by thermal velocities, focusing on the halo mass function and radial density profiles of haloes.

\section{Thermal velocities and initial conditions}\label{ICs}
This section explores the impact of thermal velocities in the N-body initial conditions. First, we introduce the WDM model chosen for our analysis. Then, we describe how to generate N-body initial conditions for such a model. Finally, we measure velocity power spectra from simulations which include thermal velocities and compare the results with simulations which ignore thermal velocities. 
\subsection{The model}

We choose for our analysis the so-called thermal WDM scenario. In this model, a non-standard fermionic species with mass $m_\mathrm{WDM}$ is in thermal contact with the primordial bath, decouples when still relativistic and enters the non-relativistic regime deep into the radiation-dominated period \cite{Bode:2000gq,Hansen:2001zv}. Since this component freezes-out when relativistic and in thermal contact, the functional form of the momentum distribution remains Fermi-Dirac after decoupling,
\begin{equation}
f_\mathrm{WDM}(p) = \frac{1}{e^{p/T_\mathrm{WDM}}+1}.
\end{equation}
The temperature redshifts as $T_\mathrm{WDM}\propto (1+z)$ and its present value can be related to the present neutrino temperature $T^0_\nu$ as \cite{Hansen:2001zv},
\[
T^0_\mathrm{WDM} = \left(\frac{\Omega^0_\mathrm{WDM}  \,h^2} {{m_\mathrm{WDM}}/{94\,\text{eV}}}\right)^{1/3} T^0_\nu, 
\] 
where $\Omega^0_\mathrm{WDM}$ is the present day fractional contribution of WDM in units of the total (critical) density of the Universe $\rho^0_c \equiv 3H_0^2 /8\pi G$ and $h \equiv H_0 / (100$ km/s/Mpc$)$ is the dimensionless value of the Hubble constant. We make the extra assumption that all the DM sector is in form of WDM, i.e. $\Omega^0_\mathrm{WDM} = \Omega^0_\mathrm{m} \sim 0.3$, where $\Omega^0_\mathrm{m}$ is the present day total contribution of the DM in units of the critical density. When WDM particles become non-relativistic, their thermal velocities remain a Fermi-Dirac distribution with rms velocity of the form \cite{{Bode:2000gq}},
\begin{equation}
\sigma_\mathrm{therm} \equiv \sqrt{\left< v^2_\mathrm{therm}\right>} = 0.0429 \left( \frac{\Omega^0_\mathrm{WDM}}{0.3} \right)^{1/3} \left( \frac{h}{0.7} \right)^{2/3} \left( \frac{\text{keV}}{m_\mathrm{WDM}}\right)^{4/3} (1+z) \,\,\, \text{km/s}.
\label{eq:rmsthermalvel}
\end{equation}

For the model discussed in this section, a parametrisation of the linear power spectrum of density fluctuations exists \cite{Bode:2000gq}, which we will use throughout. This parametrisation can be written in terms of a transfer function relative to CDM, $T^2(k) \equiv P^{\mathrm{WDM}}(k)/P^{\mathrm{CDM}}(k)$:
\begin{equation}
T(k) = \left(1 + \left(\alpha k\right)^{2 \nu}\right)^{-5 / \nu},
\label{eq:fittingformula}
\end{equation}
where 

\begin{equation}
\alpha = a  \left(\frac{\Omega^0_\mathrm{WDM}}{0.25}\right)^b  \left(\frac{h}{0.7}\right)^c  \left(\frac{m_\mathrm{WDM}}{\text{keV}}\right)^d,
\end{equation}
and 
\[
a = 0.049, \quad b = 0.11, \quad c = 1.22, \quad d = - 1.11, \quad \nu = 1.12,
\]
as computed in \cite{Viel:2005qj}. More accurate power spectra for more general non-cold DM models can be generated using Boltzmann codes such as {\sc class} \cite{2011arXiv1104.2932L,2011JCAP...09..032L}.

The characteristic scale at which the cut-off in the WDM power spectra occurs  can be defined through the half-mode wavenumber $k_\mathrm{hm}$. $k_\mathrm{hm}$ is the wavenumber at which the transfer function $T(k)$ falls to $50\%$. Given the parametrisation in eq. (\ref{eq:fittingformula}) for $T(k)$, we have 
\begin{equation}
k_\mathrm{hm} = \frac{1}{\alpha} \left(2^{\nu/5} -1 \right)^{1/2\nu}.
\label{eq:halfmodescale}
\end{equation} 

\subsection{The mass of the WDM candidates}
Thermal WDM with a particle mass around a keV has a free-streaming characteristic length of the order of dwarf galaxy scales, so candidates with such masses can solve the missing satellite problem \cite{Bode:2000gq,Colin:2000dn,Hansen:2001zv,Viel:2005qj}. The strongest constraints on the free-streaming length of DM particles come from the Lyman-$\alpha$ forest. The Lyman-$\alpha$ forest is a series of absorption lines in the spectra of distant quasars due to the inhomogeneous distribution of the intergalactic medium (IGM). These lines  can be used as a measure of the matter power spectrum on scales $0.5 \, h^{-1}\text{Mpc}<\lambda<100$ $h^{-1}$Mpc over a range of redshifts $2<z<6$ \cite{Viel:2005qj,Viel:2013apy,Irsic:2017ixq}. The current constraints on thermal WDM using the Lyman-$\alpha$ forest suggest a mass\footnote{A recent analysis pushes the limit to slightly higher masses $m_\mathrm{WDM} \gtrsim 3.5$ keV \cite{Irsic:2017ixq}.} of $m_\mathrm{WDM} \gtrsim 3.3$ keV \cite{Viel:2013apy}. However, the Lyman-$\alpha$ forest does not probe directly the DM power spectrum, but rather that of the neutral hydrogen. The neutral hydrogen gas does not necessary follow the distribution of DM, since the process of reionisation heats the hydrogen, preventing it from clustering at small scales \cite{Gnedin:1997td}. To translate the results from the Lyman-$\alpha$ forest to constraints on the DM power spectrum, a crucial parameter is the temperature of the IGM. However, the IGM temperature is not precisely known, so some approximations are made in the literature, depending on which the constraints on WDM particle mass can vary. Indeed, the $m_\mathrm{WDM} \gtrsim 3.3$ keV mass bound can be relaxed to $m_\mathrm{WDM}  \gtrsim 2$ keV by changing the assumptions about the IGM thermal history \cite{Viel:2013apy} (see also \cite{Garzilli:2015iwa}, in which the authors have re-analysed the Lyman-$\alpha$ forest spectra considered in \cite{Viel:2013apy}). Motivated by the above discussion, whilst the assumptions about the thermal history of the IGM remain controversial, we choose three benchmark values for the WDM particle mass in our analysis: $m_\mathrm{WDM}=2$ keV,  $m_\mathrm{WDM}=3.3$ keV  and $m_\mathrm{WDM} = 7$ keV (which will be our coldest WDM candidate).

\begin{figure}
\centering
\subfigure[][]{
\includegraphics[width=.67\textwidth]%
{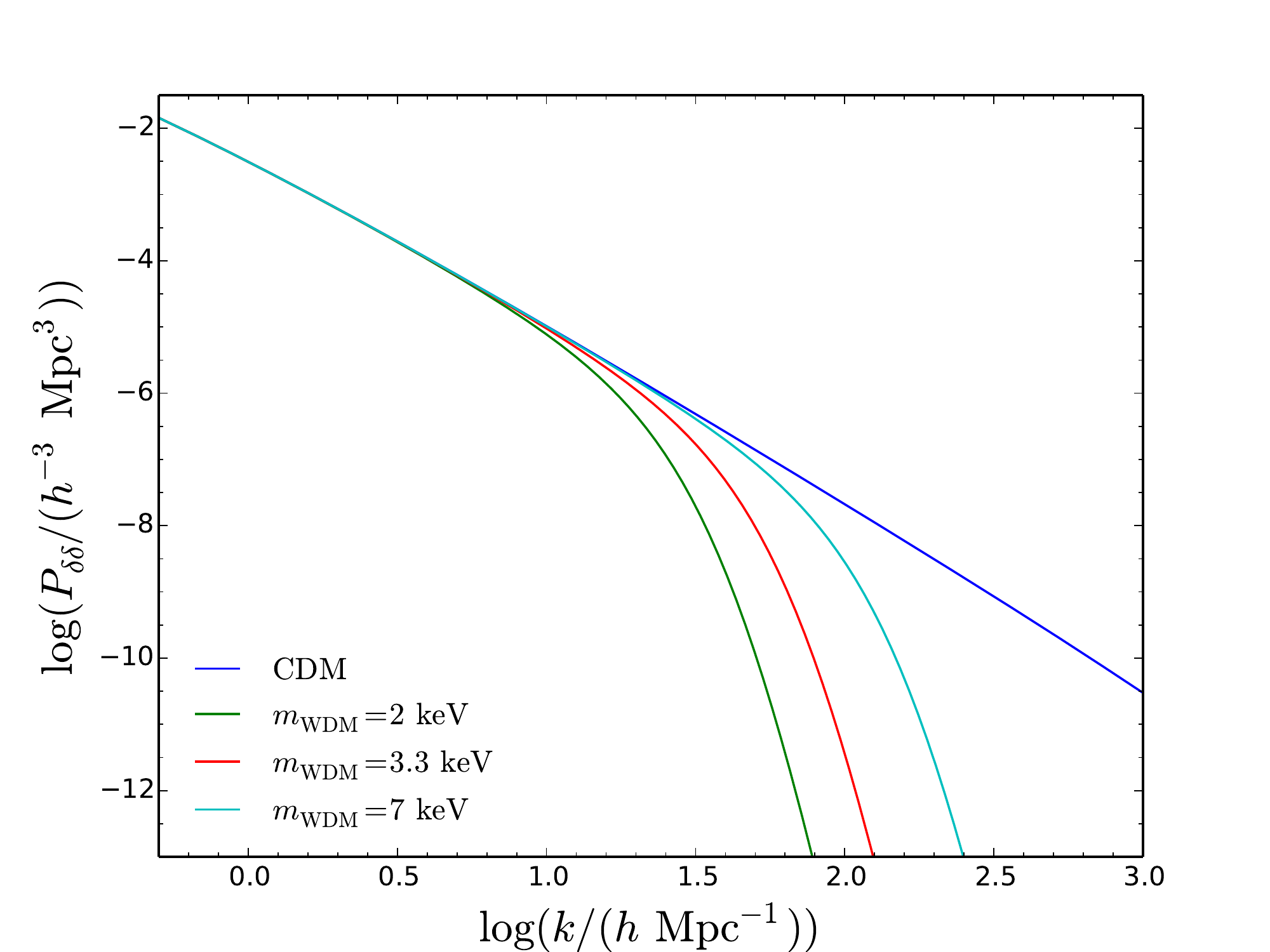}\label{fig:powerspecCDMWDMa}
}\,
\subfigure[][]{
\includegraphics[width=.67\textwidth]%
{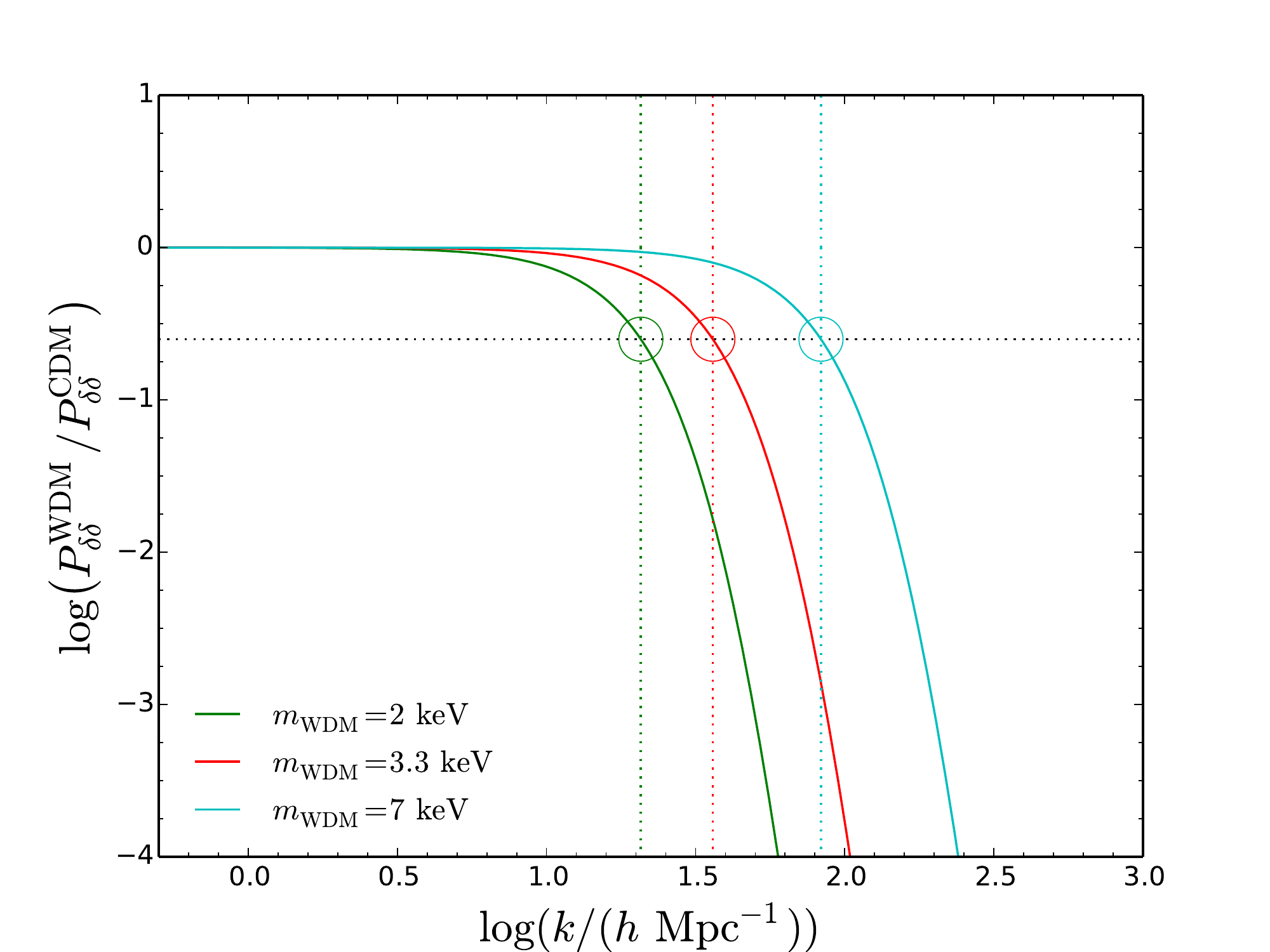}\label{fig:halfmodeks}
}
\caption{(a) Linear theory matter power spectra at redshift $z=199$ for CDM and three WDM models with masses $m_\mathrm{WDM} = 2, 3.3, 7$ keV as labelled. (b) Squared transfer functions $T^2(k) \equiv P^\mathrm{WDM}/P^\mathrm{CDM} (k)$ for the three warm candidates. The dotted black horizontal line represents the scale at which the transfer function is suppressed by a factor of two. The intersection between the transfer function and the horizontal line defines the half-mode wavenumber $k_\mathrm{hm}$.  The WDM spectra are generated using the parametrisation in eq. (\ref{eq:fittingformula}).}
\label{fig:matterspectrumz199classnewonlyCLASS}
\end{figure}

In figure \ref{fig:powerspecCDMWDMa}, we plot the linear matter power spectra  using the parametrisation in eq. (\ref{eq:fittingformula}) for the three WDM masses chosen. The CDM power spectrum is also shown for comparison\footnote{The CDM power spectrum is generated using the code {\sc class} \cite{2011arXiv1104.2932L}. The values of the cosmological parameters are given below. The DM contribution is  $\Omega^0_\mathrm{m} h^2 = 0.120$, the baryonic contribution is $\Omega_\mathrm{b} h^2 = 0.023$, the dimensionless Hubble constant is $h = 0.6726$, the spectral index of the primordial power spectrum is $n_s =  0.9652$ and the linear rms density fluctuation in a sphere of radius $8$ $h^{-1}$Mpc at $z=0$ is $\sigma_8 = 0.81$.}. In figure \ref{fig:halfmodeks} we plot the square of the transfer function $T(k)$, together with the half-mode wavenumber $k_\mathrm{hm}$ (see eq. (\ref{eq:halfmodescale})) for each of the WDM candidates. 

\subsection{The simulations}
In order to take into account thermal velocities in the N-body simulations, the initial conditions (ICs) for running cosmological simulations have to be modified accordingly. We use the numerical code 2LPTic \cite{Crocce:2006ve}, which provides ICs based on second-order Lagrangian perturbation theory.  2LPTic can generate the ICs for thermal WDM models, taking as input the thermal WDM mass $m_\mathrm{WDM}$ and computing the corresponding $T(k)$ from eq. (\ref{eq:fittingformula}). The code also contains a module for adding thermal velocities, following the approximation used in \cite{Boyarsky:2008xj,Colin:2007bk,Klypin:1992sf,Brandbyge:2008rv,Viel:2010bn,2013JCAP...03..019V,2012MNRAS.421...50V,2013MNRAS.428..882M,2013MNRAS.430.2346S,Paduroiu:2015jfa}. The thermal velocities for simulation particles are randomly picked such that their magnitudes obey a Fermi-Dirac distribution with a dispersion given by eq. (\ref{eq:rmsthermalvel}), while the directions satisfy a uniform distribution in the $4\pi$ solid angle. The addition of thermal velocities to peculiar velocities is done using a simple velocity addition law.

We run 2LPTic for the three WDM candidate masses mentioned above, $m_\mathrm{WDM}=2,3.3,7$ keV. For each mass, we consider two types of WDM simulations, (i) WDM simulations without thermal velocities (which we call $\mathrm{WDM}${-}$\mathrm{novth}$) and (ii) WDM simulations which take into account thermal velocities in the ICs (referred to as $\mathrm{WDM}${-}$\mathrm{vth}$). Our analysis consists of generating ICs for these two types of simulations, varying the number of simulation particles $N$ and the length of the box $L$. For each pair $\{N,L\}$, the Nyquist frequency of a simulation is $k_{Ny} \equiv \pi (N^{1/3}/L)$ (this specifies the value up to which we can trust the $P(k)$ estimate using an FFT and the scale down to which a clustering signal can be imposed due to the mean inter particle separation). 

\begin{table}
\centering
\footnotesize
\begin{tabular}{ccccccccc}
\toprule
$m_\mathrm{WDM}$&$L$ &  $N$&$k_\mathrm{Ny}$  &$\mathrm{WDM}${-}$\mathrm{novth}$ & $\mathrm{WDM}${-}$\mathrm{vth}$ & $z_\mathrm{ini}$ & Evolved&$\epsilon$ \\
(keV)& ($h^{-1}$Mpc)               &  & ($h$ Mpc$^{-1}$)  & &  &  & & ($h^{-1}$kpc) \\

\midrule
\hphantom{0}$2$ 		&    $\hphantom{0}2$ &$512^3$ &$803.8$& \checkmark&\checkmark & $199$  &$\times$& -\\
\hphantom{0}$2$    & $25$&$512^3$ &$\hphantom{0}64.3$& \checkmark&\checkmark & $199$&\checkmark & 1.22\\
$3.3$    & $\hphantom{0}2$&$\hphantom{0}64^3$&$100.5$ & \checkmark&\checkmark & $199$&$\times$& -\\
$3.3$    & $\hphantom{0}2$&$128^3$ &$201.0$& \checkmark&\checkmark & $199$&$\times$& -\\
$3.3$    & $\hphantom{0}2$&$256^3$&$401.9$ & \checkmark&\checkmark & $199$&$\times$& -\\
$3.3$    & $\hphantom{0}2$&$512^3$&$803.8$ & \checkmark&\checkmark & $199$&$\times$& -\\
$3.3$    & $10$&$512^3$&$160.8$ & \checkmark&\checkmark & $199$&$\times$& -\\
$3.3$    & $25$&$512^3$&$\hphantom{0}64.3$ & \checkmark&\checkmark & $199$ &\checkmark & 1.22\\
$3.3$    & $50$&$512^3$&$\hphantom{0}32.2$ & \checkmark&\checkmark & $199$&$\times$& -\\
$3.3$    & $25$&$512^3$&$\hphantom{0}64.3$ & \checkmark&\checkmark & $\hphantom{0}39$ &\checkmark &1.22\\
\hphantom{0}$7$ 	   &    $\hphantom{0}2$ &$512^3$&$803.8$ & \checkmark&\checkmark & $199$ &$\times$& -\\
\hphantom{0}$7$    & $12$&$512^3$&$134.0$ & \checkmark&\checkmark & $199$ &\checkmark & 0.58\\
$\hphantom{0}7$    & $25$&$512^3$&$\hphantom{0}64.3$ & \checkmark&\checkmark & $199$ &\checkmark &1.22\\
\bottomrule
\end{tabular}
\caption{Summary of the simulations performed. $m_\mathrm{WDM}$ is the physical mass of the WDM candidate. $L$ and $N$ are the simulation box length and the number of simulation particles respectively. $k_\mathrm{Ny}$ is the Nyquist frequency.  $\mathrm{WDM}${-}$\mathrm{novth}$ refers to simulations which ignore thermal velocities, while $\mathrm{WDM}${-}$\mathrm{vth}$ refers to simulations which include thermal velocities. $z_\mathrm{ini}$ is the redshift at which the ICs are generated. The checkmarks (\checkmark) in the column ``Evolved'' indicate the simulations which have been evolved up to $z=0$. For the ICs which have been evolved, we also show the gravitational softening length $\epsilon$.}
\label{table:allsim}
\end{table}
 
The ICs are generated at the initial redshift $z_\mathrm{ini} =199$, when the thermal velocity dispersions are, see eq. (\ref{eq:rmsthermalvel}),
\[
\begin{split}
&\sigma_\mathrm{therm} \simeq 3.4 \, \mathrm{km/s} \quad \mathrm{for}\,\,\, m_\mathrm{WDM} = 2 \, \mathrm{keV},\\
&\sigma_\mathrm{therm} \simeq 1.7 \, \mathrm{km/s} \quad \mathrm{for}\,\,\, m_\mathrm{WDM} = 3.3 \, \mathrm{keV},\\
&\sigma_\mathrm{therm} \simeq 0.6\, \mathrm{km/s} \quad \mathrm{for}\,\,\, m_\mathrm{WDM} = 7 \, \mathrm{keV}.\\
\end{split}
\]
These are non-negligible with respect to the peculiar velocities, which are of the order of $v_\mathrm{pec} \sim 10$ km/s at $z=199$. We have run an additional simulation with a  lower initial redshift $z_\mathrm{ini}=39$ to test the impact of the initial redshift on the results; we postpone a discussion of this to Section 3. In Section 3 we have also evolved some of the ICs up to $z=0$, using the publicly available tree-PM code Gadget-2 \cite{Springel:2005mi}. The gravitational softening length $\epsilon$ is set to be $1/40$-th of the mean interparticle separation, $L/N^{1/3}$. Further details of our simulations are listed in table \ref{table:allsim}.
\subsection{Velocity power spectra measurement}
In order to quantify the effects of thermal velocities in $\mathrm{WDM}${-}$\mathrm{vth}$, we measure the matter and velocity power spectra from the simulations. However, because in the initial conditions thermal velocities are implemented only in the velocity sector, by default the matter perturbations in the initial conditions are not affected. Instead, the effect of thermal velocities on the matter distribution only becomes apparent after at the first time step and through the subsequent evolution. As in this section we are interested only in the ICs, we shall focus on the velocity power spectra extracted from the ICs. The velocity (divergence) power spectra are defined as \cite{Pueblas:2008uv}, 
\begin{equation}\label{eq:normvelspec}
P_{\theta\theta}(z, k)=\frac{1}{(a(z)H(z) f(z))^2}\left<\theta^2_{\vec{k}} (z)\right>,
\end{equation}
where $\theta_{\vec{k}} (z)$ is the Fourier transform of the velocity divergence $\theta(\vec{x},z) \equiv \vec{\nabla} \cdot \vec{v}(\vec{x},z)$ and $f(z) = d \ln D_+/ d \ln a$, with $D_+$ linear growth factor and $a$ scale factor. Remember that the matter power spectrum is 
\begin{equation}
P_{\delta\delta}(z, k)\equiv \left<\delta^2_{\vec{k}} (z)\right>, 
\end{equation}
where $\delta \equiv \delta \rho /\bar{\rho}$ is the matter overdensity. So the normalisation in eq. (\ref{eq:normvelspec}) is useful since in linear regime  $\theta = -aHf\delta$ and the velocity power spectrum is equal to the matter power spectrum, $P_{\theta\theta} = P_{\delta\delta}$. For this reason, we can compare the velocity power spectra measured from simulations directly with the matter power spectra specified by the fitting formula in eq. (\ref{eq:fittingformula}). 

Extracting the velocity field from a N-body particle distribution is more challenging than constructing the matter density field, due to the non-additivity of the velocities\footnote{The non-additivity of the velocities refers to the fact that the net (center-of-mass) velocity of a set of particles is not the sum of the particle velocities $\vec{v}_i$ in the set, but is instead the mass-weighted average, $\vec{v}_\mathrm{net}=\sum_i w_i \vec{v}_i$, with $w_i = m_i/M$, where $m_i$ is the mass of the $i$-th particle and $M=\sum_i m_i$.}. Indeed, as pointed out in \cite{Pueblas:2008uv}, applying the standard methods used to estimate matter power spectra will not always give the expected results when we consider velocity power spectra. This is mainly because the standard methods usually use a grid-based scheme, in which the simulation box is covered by a grid. At each vertex a value for the matter density (or momentum density), is calculated using some mass assignment scheme (e.g. nearest grid point, cloud in cell, triangular shaped cloud), in the form of a weighted sum of the masses (or momenta) of the particles near each grid vertex (see e.g. \cite{1988csup.book.....H}). Usually, the velocity field is obtained by averaging the particle velocities near each corresponding grid vertex.  Thus, in order to assign a velocity field value at each vertex, we need first to assign the momenta and then divide it by the mass density at the same vertex. The limitation of the method is clearly due to the $0/0$ divergence caused by empty grid cells, in which there could be no particles at all and then the momenta and mass density both add to zero. 

As suggested in \cite{Pueblas:2008uv}, an alternative way to measure the velocity power spectra is to use the Delaunay Tessellation methodology \cite{Bernardeau:1995en}. The Delaunay Tessellation for a set of points $p$ distributed in a 3D space is the set of tetrahedrons (where each vertex is a point of the set) whose circumscribed sphere contains no points other than the four that generate the tetrahedron (and that lie on its surface). This construction is unique. The velocity at the general coordinate $\vec{r}=(x,y,z)$ inside a tetrahedron can then be inferred using a linear interpolation between the velocity values at the 4 vertices of the tetrahedron. In this case when we split the box into cells, we avoid the singularities, since the cell will be contained in at least one tetrahedron. The velocity associated with each grid vertex is computed as the average (in volume) given by all Delaunay tetrahedra contained in the grid cell corresponding to such a point \cite{Pueblas:2008uv}. In this way, the only way to have zero velocity field in a given grid point is a zero mean among all the tetrahedrons contained within the corresponding cell. Throughout we will use the publicly available DTFE code to estimate velocity power spectra, which implements the Delaunay tessellation approach \cite{2011ascl.soft05003C}.

\subsection{Results and discussion}

\begin{figure}[]
\centering
\subfigure[][$N=512^3$, $m_\mathrm{WDM} = 3.3$ keV. Varying $L$. ]{
\includegraphics[scale=.377]%
{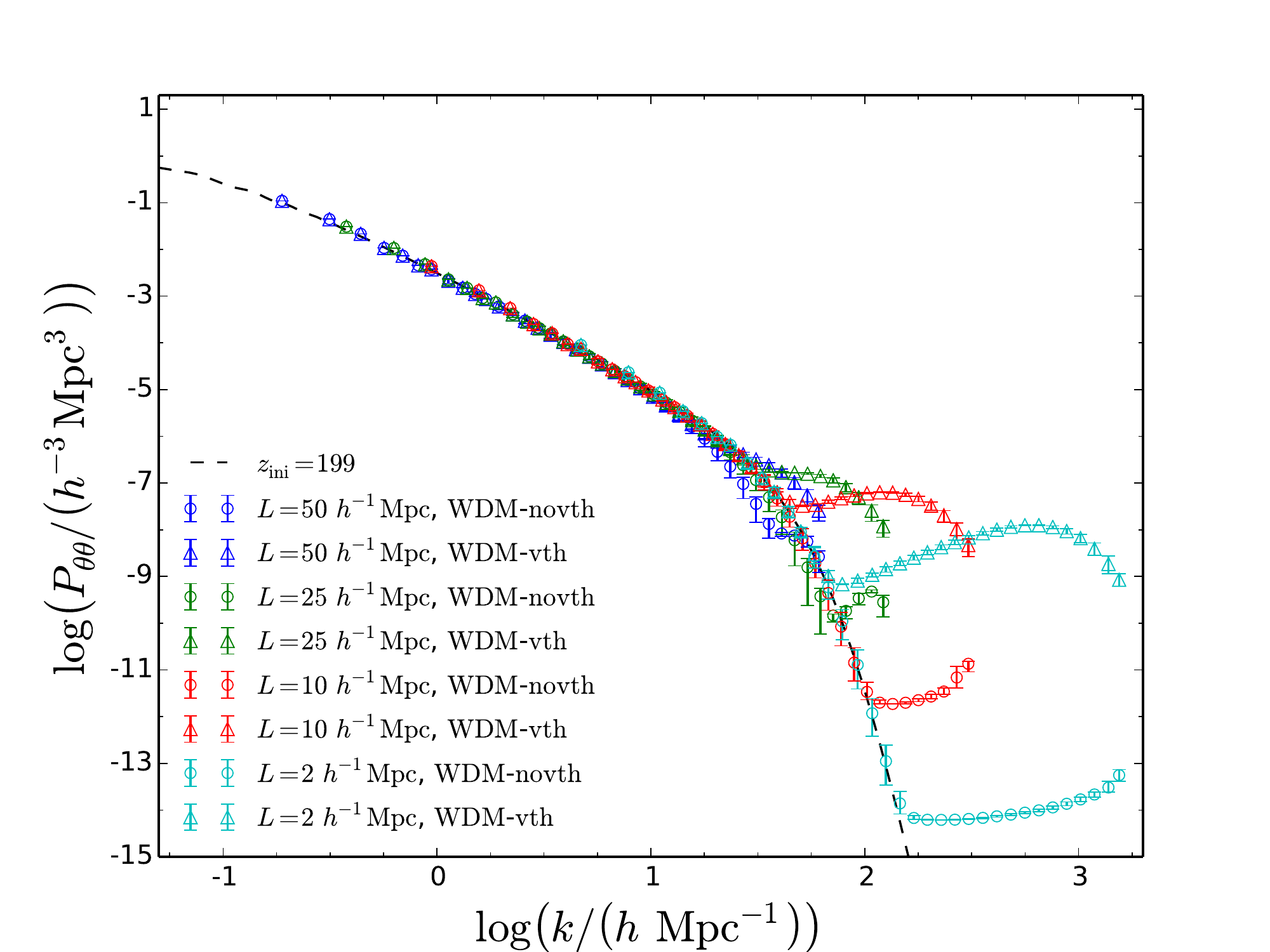}\label{fig:boxsizeics}}\\[-1.45ex] 
\subfigure[][$L = 2$ $h^{-1}$Mpc, $m_\mathrm{WDM} = 3.3$ keV. Varying $N$. ] {
\includegraphics[scale=.377]%
{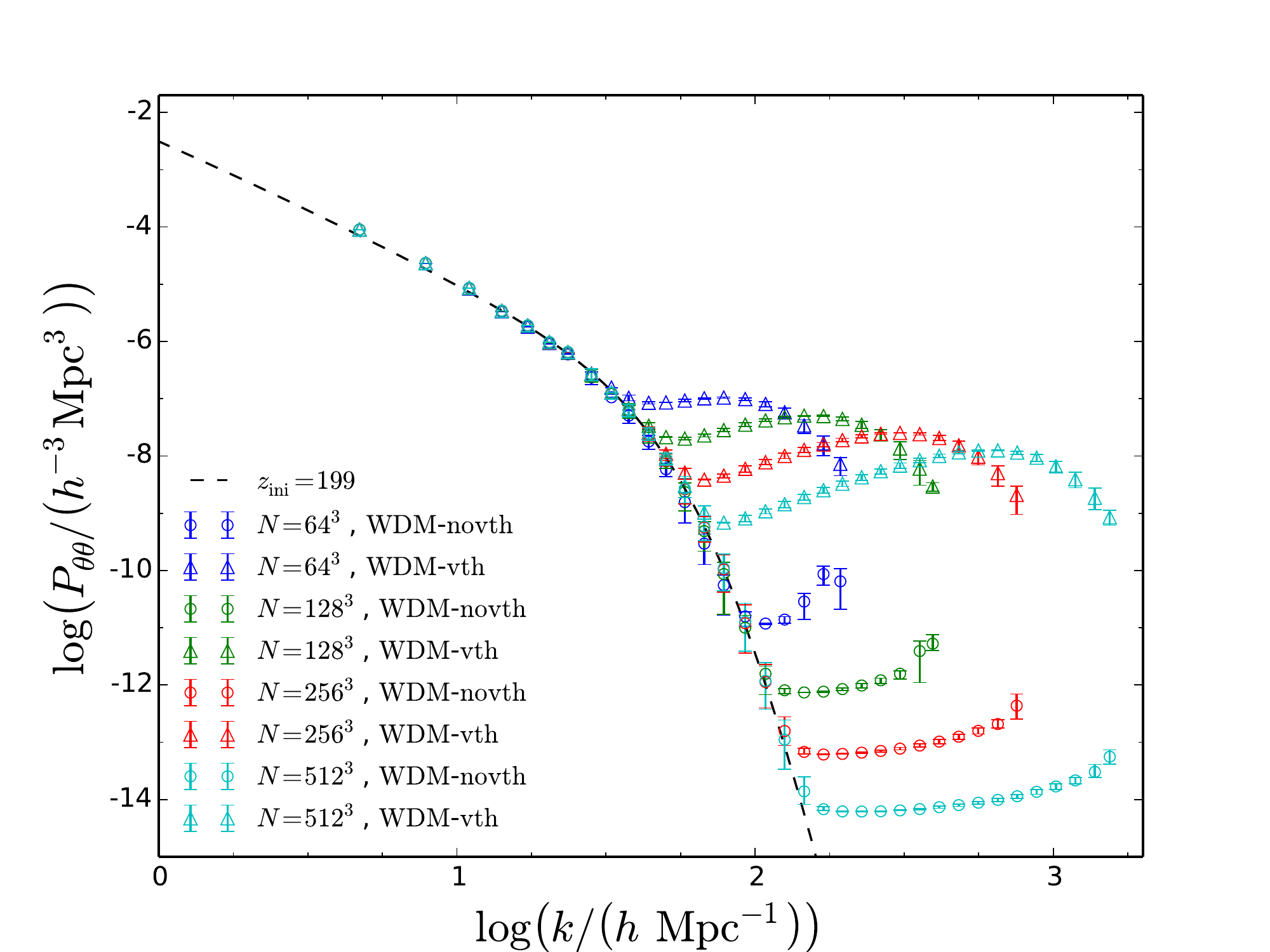}\label{fig:partics}}\\[-1.45ex]
\subfigure[][$N=512^3$, $L = 2$ $h^{-1}$Mpc. Varying $m_\mathrm{WDM}$. ]{
\includegraphics[scale=.377]%
{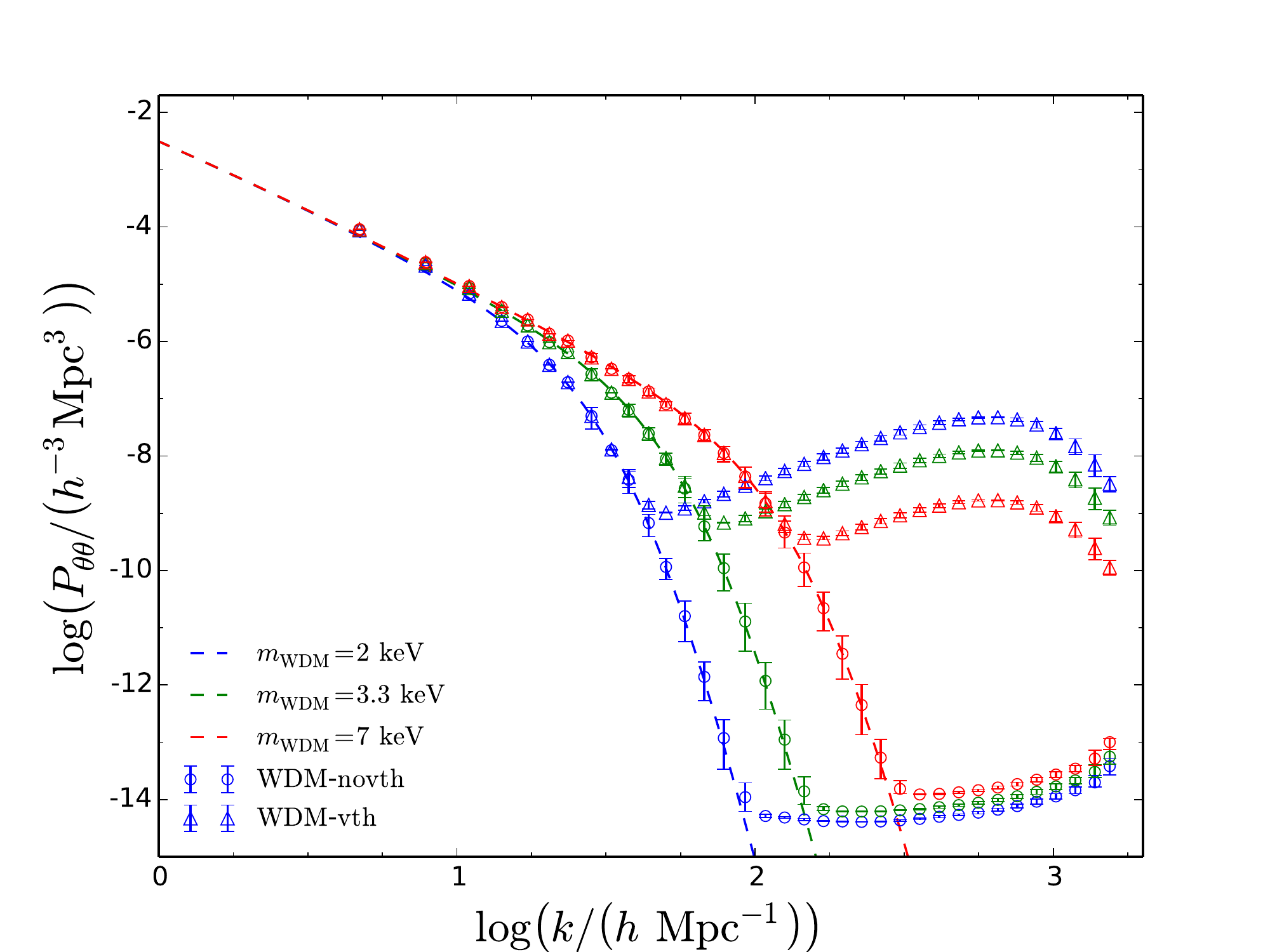}\label{fig:massdiff1}}\\[-2.5ex]

\caption{Velocity power spectra measured from ICs for $\mathrm{WDM}${-}$\mathrm{vth}$ (triangles) and $\mathrm{WDM}${-}$\mathrm{novth}$ (dots). The different panels show how the spectra change when varying (a) the box length $L$, (b) the number of particles $N$ and (c) the mass of WDM $m_\mathrm{WDM}$, while fixing the other parameters. We normalise the velocity power spectra such that $P_{\theta\theta} = P_{\delta\delta}$ (see eq. (\ref{eq:normvelspec})), in this way they can be directly compared with the theoretical linear matter power spectra given by eq. (\ref{eq:fittingformula}) (represented as black dashed lines in this figure). Note that the range of scales plotted on the $x$-axis and $y$-axis in each panel is different.}
\label{fig:allboxsizeics}
\end{figure}

Figure \ref{fig:allboxsizeics} shows the velocity power spectra measured from some of the simulations in table \ref{table:allsim}. In the first panel (figure \ref{fig:boxsizeics})  we show the results for simulations with box length $L =\{50, 25, 10, 2\}$ $h^{-1}$Mpc, whilst fixing the number of particles at $N = 512^3$  for $m_\mathrm{WDM}= 3.3$ keV. On one hand, we see that all the velocity power spectra from  $\mathrm{WDM}${-}$\mathrm{novth}$  tend to agree well with the linear theory power spectra. The agreement extends to higher wavenumbers when decreasing the box size as expected since the mean interparticle separation, $L/N^{1/3}$, goes down and consequently $k_\mathrm{Ny}$ acquires a higher value. On the other hand, the velocity power spectra extracted from $\mathrm{WDM}${-}$\mathrm{vth}$ deviate strongly from $\mathrm{WDM}${-}$\mathrm{novth}$ (and from linear theory predictions), and show an upturn at small scales. These deviations are numerical artefacts since they are pushed to higher wavenumbers for smaller boxes.

In figure \ref{fig:partics}, we show the velocity power spectra for simulations with varying number of particles $N=\{64^3,128^3,256^3,512^3\}$ in a fixed box of length $L=2$ $h^{-1}$Mpc. As we can see also in this case $\mathrm{WDM}${-}$\mathrm{vth}$ deviates from $\mathrm{WDM}${-}$\mathrm{novth}$  at high wavenumbers. These deviations tend to appear at higher wavenumbers for larger $N$, confirming the previous statement that the deviations are due to numerical noise.

The thermal velocity dispersion depends on the mass of WDM (see eq. (\ref{eq:rmsthermalvel})). So we expect that the impact of thermal velocities on the ICs will vary with $m_\mathrm{WDM}$. Figure \ref{fig:massdiff1} shows how the behaviour changes by varying $m_\mathrm{WDM}$, while fixing the number of simulated particles at $N = 512^3$ and the box length at $L = 2$ $h^{-1}$Mpc.  As we can see, increasing the mass (reducing the thermal velocity dispersion) reduces the deviations due to the noise in  $\mathrm{WDM}${-}$\mathrm{vth}$ and shifts the discrepancies to higher wavenumbers. For the most massive candidate ($m_\mathrm{WDM}=7$ keV) there is agreement between $\mathrm{WDM}${-}$\mathrm{vth}$ and $\mathrm{WDM}${-}$\mathrm{novth}$ up to $k\approx 170$  $h/$Mpc, while for the lightest mass ($m_\mathrm{WDM} = 2$ keV) the power spectrum from simulations with thermal velocities starts to disagree at $k\gtrsim 40$  $h/$Mpc.  

The deviations in the simulations with thermal velocities are due to numerical noise. The numerical noise originates in the way we assign a thermal velocity vector to a simulation particle. This can be understood by the following argument. The ICs are generated starting from a homogeneous grid of simulation particles and associating a velocity vector $\vec{v}$ at each particle on the grid. Simulation particles in CDM simulations (or in WDM simulations without thermal velocities) have no thermal velocities, so only peculiar velocities contribute to the velocity vector at each grid point, $\vec{v}=\vec{v}_\mathrm{pec}$. On the other hand, in WDM simulations with thermal velocities we add a random thermal velocity vector to each simulation particle, so the net velocity at each grid point is $\vec{v}=\vec{v}_\mathrm{pec}+\vec{v}_\mathrm{therm}$. The simulation particles are initially put onto a grid to suppress shot noise effects. This is sometimes referred to as a ``quiet start''. When the particles are displaced from a grid the measured matter power spectrum agrees with the target linear theory predictions down to smaller scales than is the case if the particles have a random distribution before they are displaced (see e.g. the discussion on initial particle arrangements in \cite{Baugh:1994hb}). However, when thermal velocities are included the particles move, on average, a greater distance away from their initial grid positions after the first time step, and effectively ``lose memory'' of where they started from. Thus, the net effect on the matter power spectrum after the simulation has been evolved for one time step is almost the same as perturbing a random initial particle distribution rather than a ``quiet'' grid. Therefore, including thermal velocities in the simulation introduces more shot noise compared to that in simulations without thermal velocities. It is well known that the shot noise level is reduced when increasing the simulation resolution (i.e. increasing the number density of the simulation particles); indeed in our results the noise effect is pushed to smaller scales as the resolution improves for simulations with thermal velocities (see figures \ref{fig:boxsizeics} and \ref{fig:partics}).

Another way to reduce the noise in the ICs is to choose a lower initial redshift. This is because the impact of thermal velocities decreases linearly with redshift (see eq. (\ref{eq:rmsthermalvel})). However, we cannot make the initial redshift arbitrarily low since at very low redshifts nonlinear corrections to the power spectrum, especially at small scales, are no longer negligible. In the next section we will discuss how the noise is reduced if we start at a lower redshift.

\section{Structure formation: evolved matter and velocity spectra}
This section is devoted to a study of the nonlinear evolution of the ICs. We consider two sets of ICs, one generated at $z_\mathrm{ini}=199$ and the other at a lower redshift, $z_\mathrm{ini}=39$, which is chosen because at this redshift the thermal velocities are negligible with respect to peculiar velocities, so we expect that the numerical noise does not affect the ICs.  All simulations are performed in a cubic box of length $L=25$ $h^{-1}$Mpc using $N=512^3$ particles. We choose this pair of $\{N,L\}$ in our simulations since we want to resolve the structures at scales near the half-mode wavenumber (see eq. (\ref{eq:halfmodescale})) of our warmer candidates\footnote{For $m_\mathrm{WDM} = 2$ keV and $3.3$ keV, the half-mode wavenumbers are within the scales probed accurately by a simulation with $\{N=512^3,L=25 h^{-1}\mathrm{Mpc}\}$. However, for $m_\mathrm{WDM} = 7$ keV the half-mode wavenumber is larger than the Nyquist frequency of the simulation, $k_\mathrm{Ny}\sim 64$ $h/$Mpc. In Section 4 (when analysing halo properties) we will show some results from a high-resolution simulation, which resolves the half-mode scale for WDM with $m_\mathrm{WDM} = 7$ keV.} ($m_\mathrm{WDM} = 2$ keV and $3.3$ keV). 

We evolve the ICs up to $z = 0$, showing how the noise in the ICs affects the matter and velocity power spectra at intermediate redshifts.

\subsection{Results for initial conditions generated at $z_\mathrm{ini}=199$}
\begin{figure}
\centering
 \vspace{-5.7\baselineskip}
\subfigure   
   {\includegraphics[width=.72\textwidth]{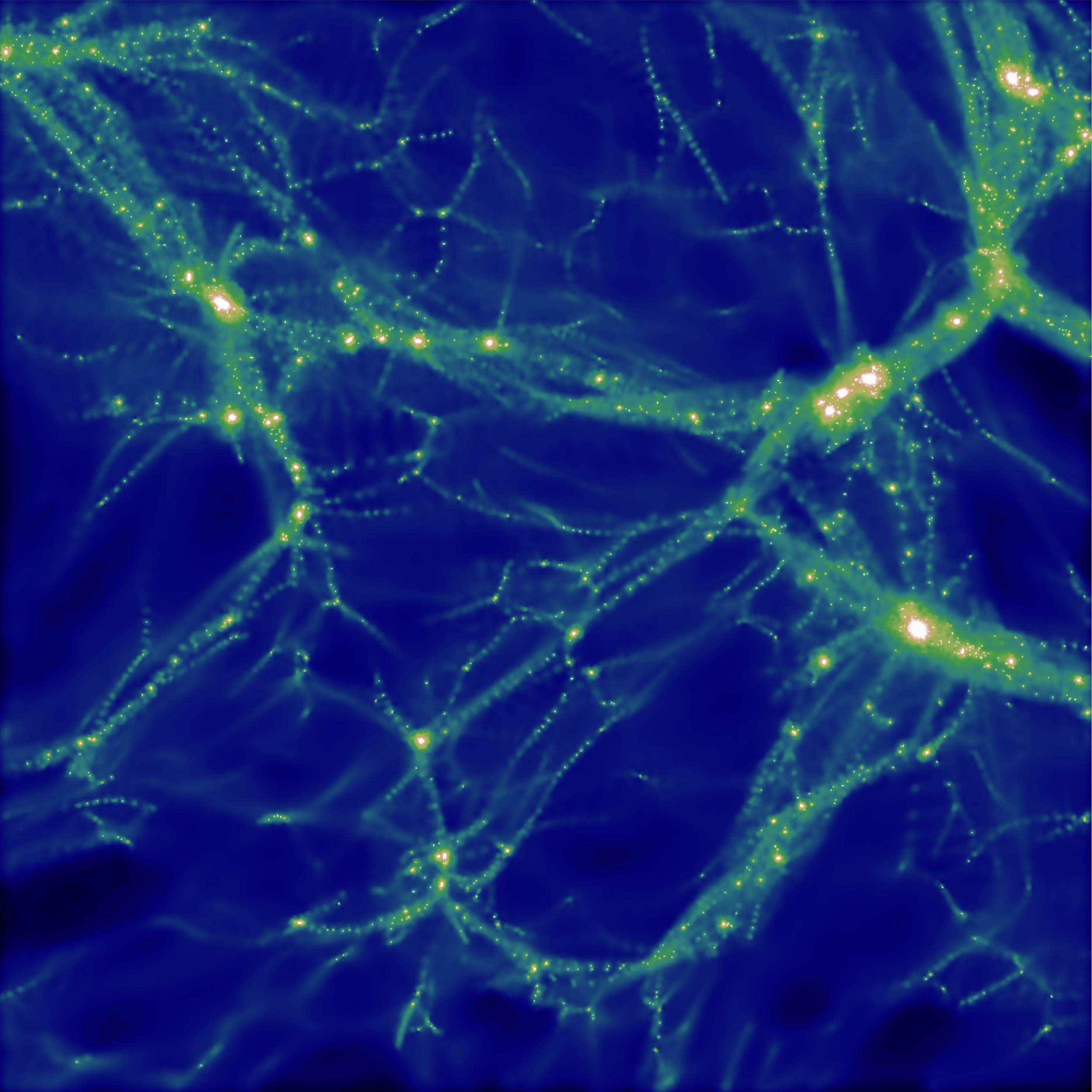}}\\\vspace{-0.7\baselineskip}
\subfigure
   {\includegraphics[width=.72\textwidth]{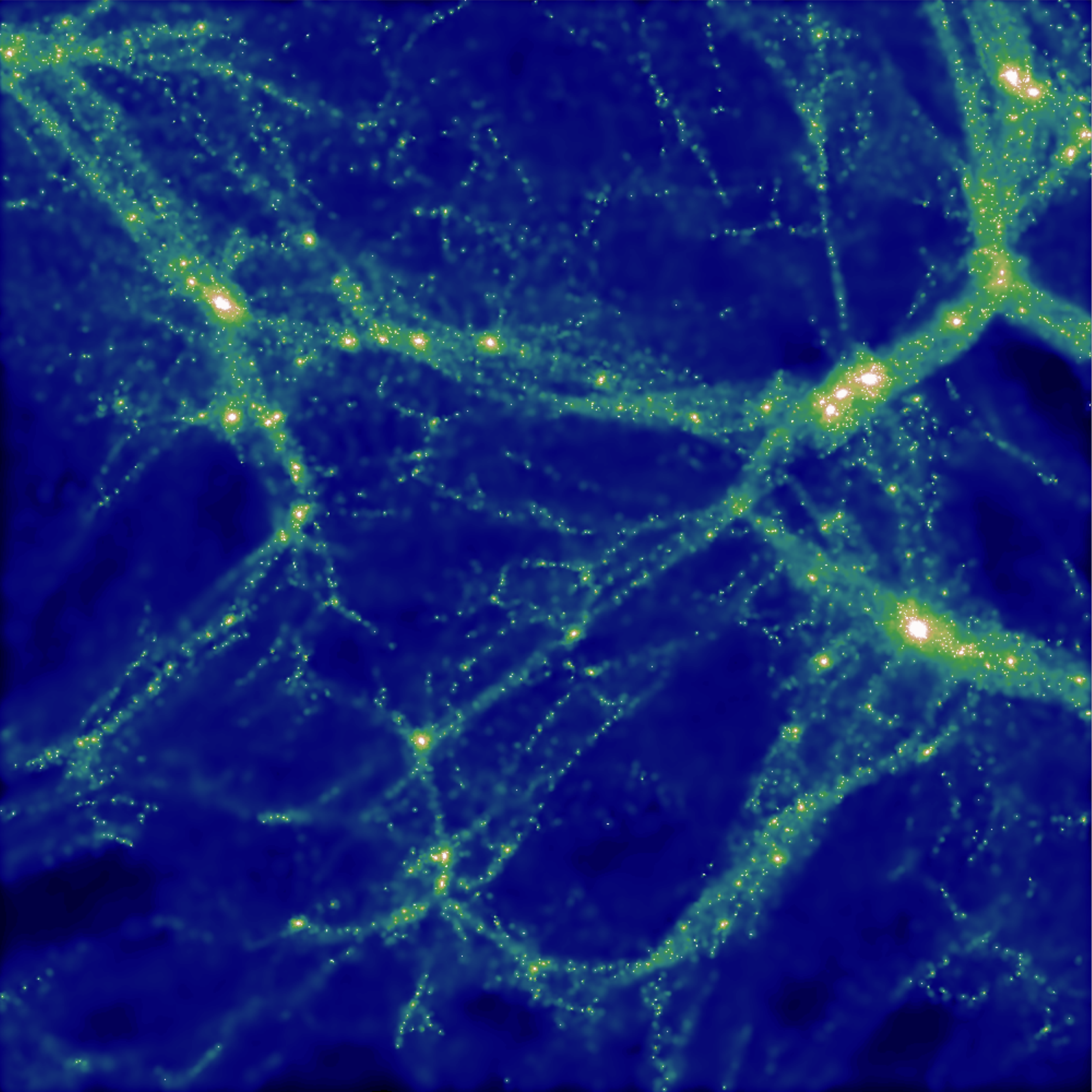}} \\ 
\caption{Logarithm of the projected mass density field of a region of $(12.5 \times 12.5 \times 2)$ $h^{-3}$Mpc$^3$ at $z=0$ considering a WDM candidate with mass $m_w = 2$ keV. The upper panel is from $\mathrm{WDM}${-}$\mathrm{novth}$, while the lower is from $\mathrm{WDM}${-}$\mathrm{vth}$. The images are generated using the code Py-SPHViewer \cite{key-SPHViewer}.}
\label{fig:snapshots}
\end{figure}

Before presenting the quantitative results of our study, let us first show two snapshots of the matter density field to qualitatively appreciate the effects introduced by including thermal velocities in the simulations. In figure \ref{fig:snapshots} we select the same region of the simulation box at $z=0$ from $\mathrm{WDM}${-}$\mathrm{novth}$ (upper panel) and from $\mathrm{WDM}${-}$\mathrm{vth}$ (lower panel) for the WDM candidate with mass $m_\mathrm{WDM}=2$ keV. It is well known that WDM simulations - due to the cut-off in the initial power spectrum - display the effects of artificial fragmentation, with regularly-spaced clumps (spurious haloes) along filaments, the distance between which reflects the initial inter-particle distance \cite{Bode:2000gq,Wang:2007he,Lovell:2013ola,Schewtschenko:2014fca,2012MNRAS.424..684S,Power:2013rpw,Power:2016usj}. Such spurious haloes are numerical artefacts and can be removed in an attempt to obtain clean, physical, halo catalogues \cite{Wang:2007he,Lovell:2013ola}. We find that this artificial effect becomes more prominent when thermal velocities are included in the simulations, due to the extra noise they introduce. In the case of $2$ keV WDM particles, we find (see Section 4 for more details) that such spurious structures dominate for haloes with mass $M<10^9h^{-1}$M$_\odot$, which are shown as white dots in figure \ref{fig:snapshots}. A quick inspection of figure \ref{fig:snapshots} confirms that simulations with thermal velocities generally show more such structures. However, the spacing of the spurious haloes is less regular than in simulations without thermal velocities. This is because the regularity of the spurious haloes reflects the grid distribution of the simulation particles in the ICs. However, when adding thermal velocities the particles move further than without thermal velocities, and ``loose memory'' of the initial regular distribution (see the discussion in Section 2). This noise effect will also be reflected in terms of enhanced matter clustering on small scales, as we shall see below.

\begin{figure}[]
\advance\leftskip-1.cm
\advance\rightskip-2cm
\subfigure[][CDM.]
{\includegraphics[width=.62\textwidth,height=7.5cm]{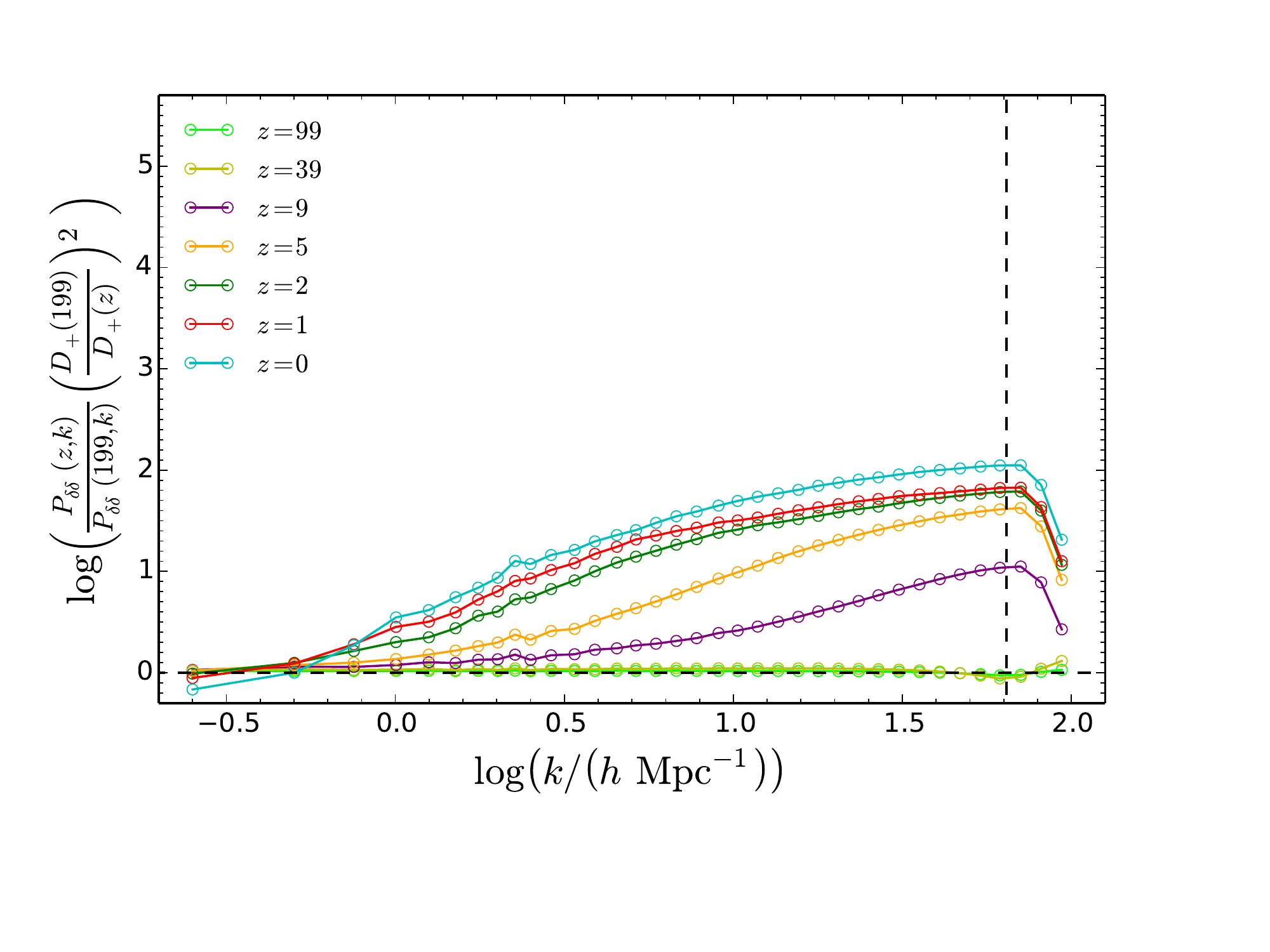}\label{fig:allmassesa}} \hspace{-3.1\baselineskip}
\subfigure[][$m_\mathrm{WDM} = 7$ keV.]
{\includegraphics[width=.62\textwidth,height=7.5cm]{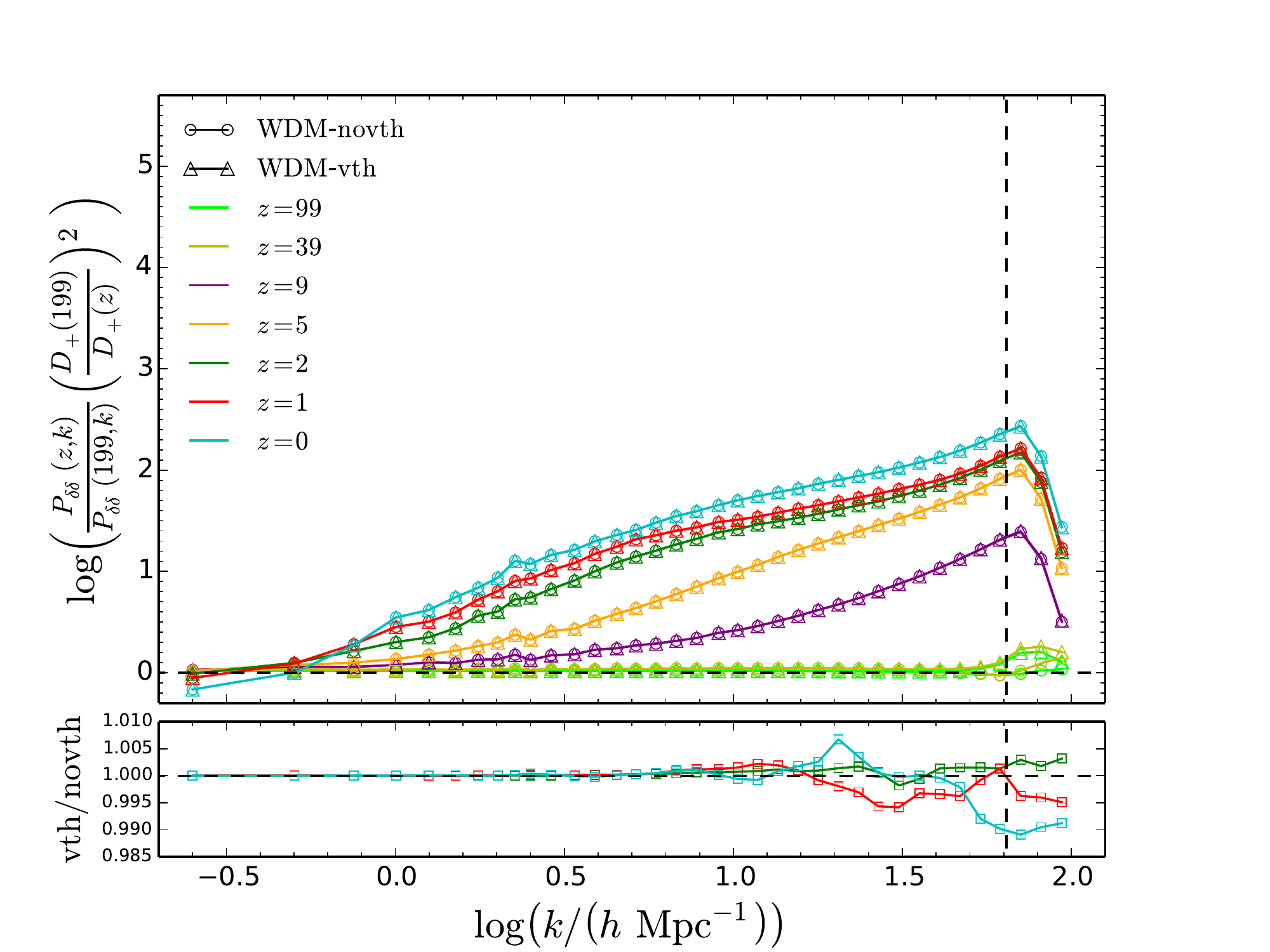}\label{fig:allmassesb}} \\
\subfigure[][$m_\mathrm{WDM} = 3.3$ keV.]
{\includegraphics[width=.62\textwidth,height=7.5cm]{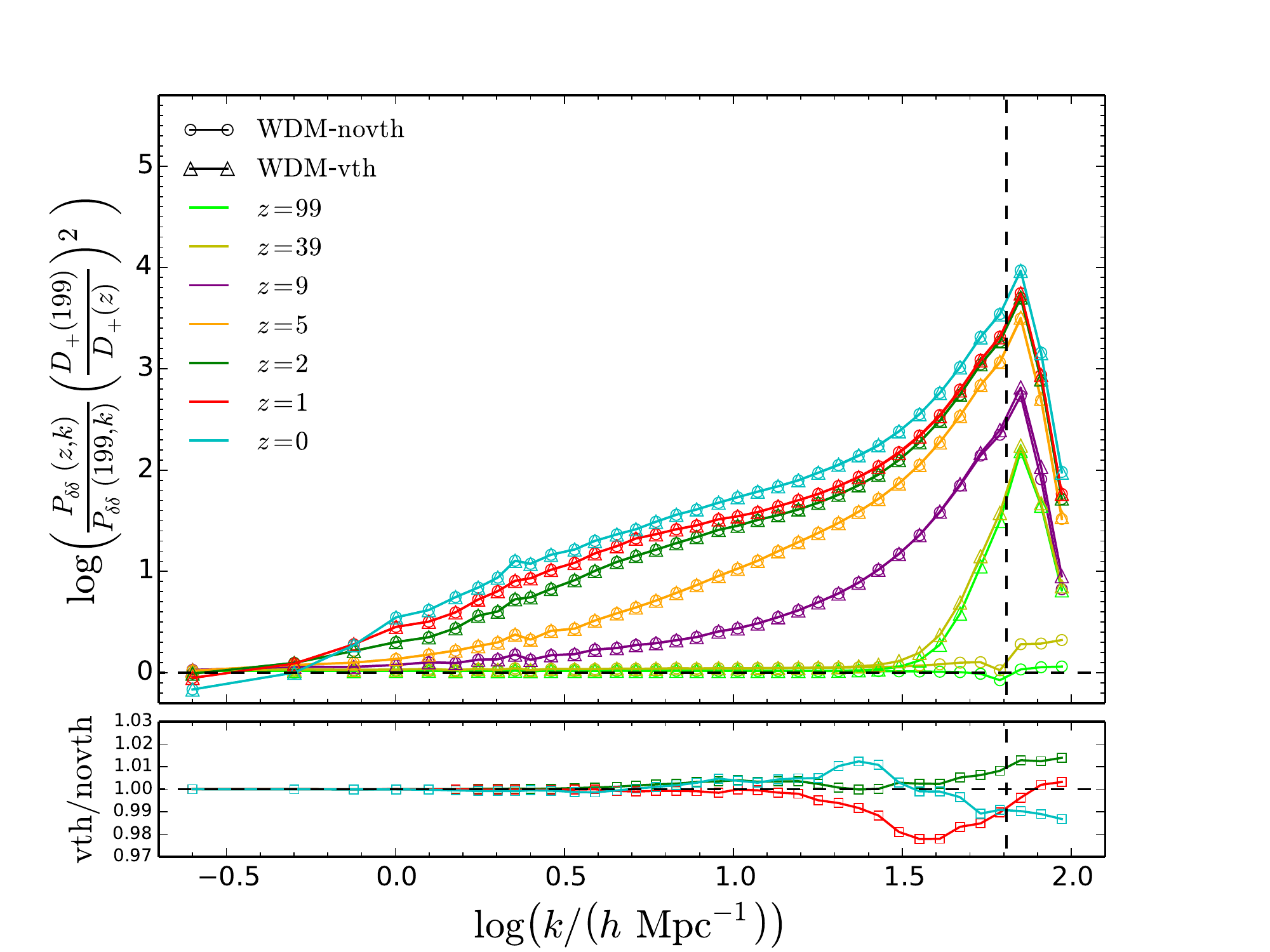}\label{fig:allmassesc}} \hspace{-3.1\baselineskip}
\subfigure[][$m_\mathrm{WDM} = 2$ keV.]
{\includegraphics[width=.62\textwidth,height=7.5cm]{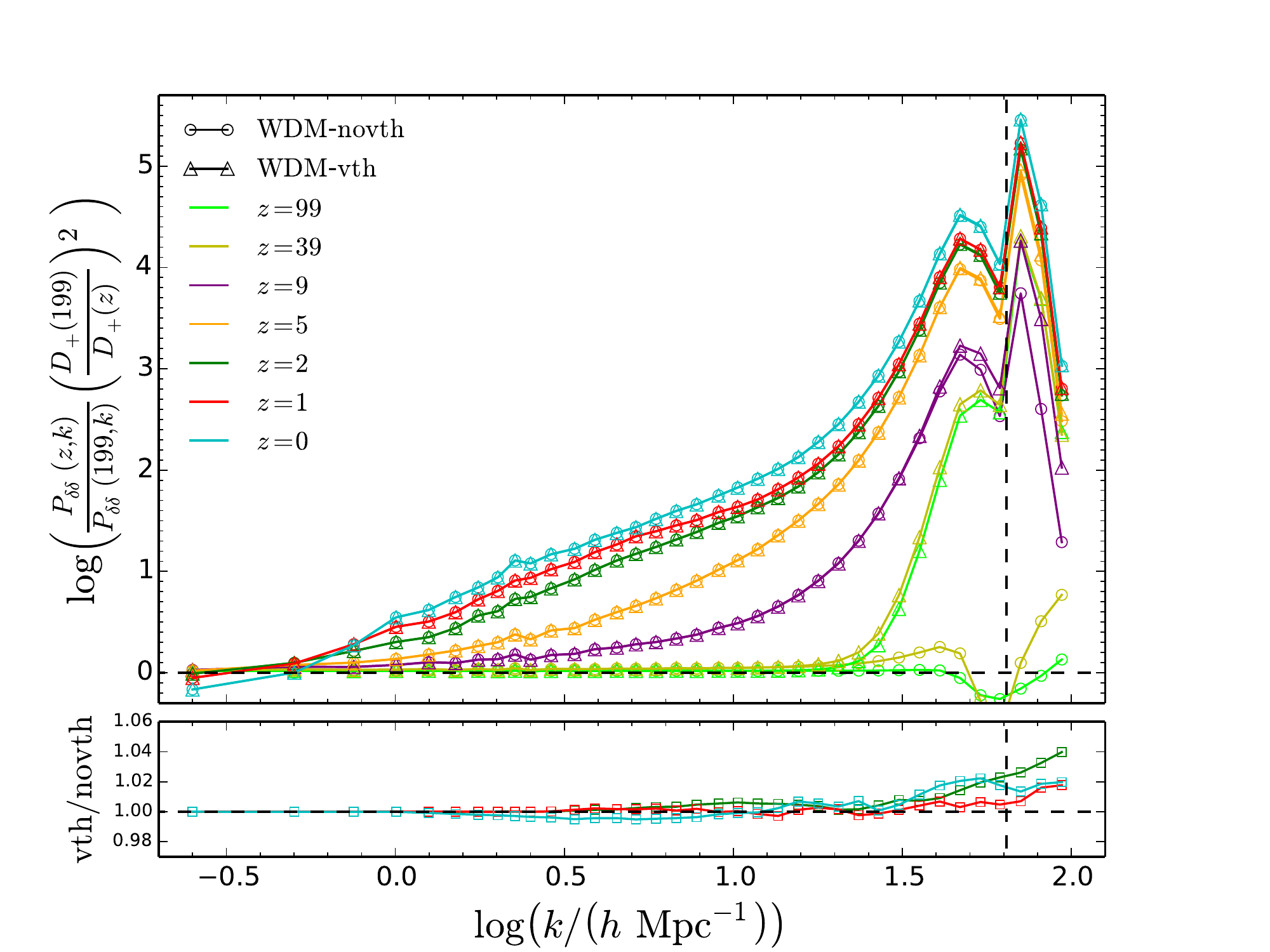}\label{fig:allmassesd}} \\   
\caption{Matter power spectra for (a) CDM and WDM with mass  (b) $m_\mathrm{WDM}=7$ keV, (c) $m_\mathrm{WDM}=3.3$ keV and (d) $m_\mathrm{WDM}=2$ keV, evolved with GADGET-2 up to redshift $z=0$. Triangles represent $\mathrm{WDM}${-}$\mathrm{vth}$, while dots are $\mathrm{WDM}${-}$\mathrm{novth}$. The power spectra are plotted as ratios respect to the initial power spectrum at $z_\mathrm{ini}=199$ scaled to take out the difference between the $\Lambda$CDM growth factor at $z = 199$ and the evolved simulation at redshift $z$. The bottom panels in figures (b-d) show the ratio between $\mathrm{WDM}${-}$\mathrm{vth}$ and $\mathrm{WDM}${-}$\mathrm{novth}$ at redshifts $z=2$ (green), $z=1$ (red), $z=0$ (cyan). The dashed vertical line shows the Nyquist frequency.}
\label{fig:allmasses}
\end{figure}

In order to quantify the effects of thermal velocities, in figure \ref{fig:allmasses} we show the matter power spectra from the evolved simulation outputs in the range of redshifts $z\in [199, 0]$. The matter power spectra are measured using the code {\sc powmes} \cite{Colombi:2008dw}. We present the evolved matter power spectra by plotting the nonlinear growth $\left(P(z)/P(199)\right) \left(D_+(199)/D_+(z)\right)^2$, where $\left(D_+(199)/D_+(z)\right)^2$ is the ratio between the $\Lambda$CDM growth factor at $z = 199$ and at redshift $z$. This rescaling highlights the changes in the shape of the power spectrum due to nonlinear growth, removing the much bigger change in the amplitude of the power spectrum due to linear growth over a large change in redshift. For comparison we have also displayed the non-linear growth for CDM matter power spectra. 

At high redshifts ($z\in[ 199,39]$), the wavenumbers probed by our simulations are well inside the linear regime, so the power spectrum ratio plotted is close to unity, which indicates that the power spectrum is evolving according to linear theory for both the CDM and the three $\mathrm{WDM}${-}$\mathrm{novth}$ simulations, see figure \ref{fig:allmasses}. However, the matter power spectra of $\mathrm{WDM}${-}$\mathrm{vth}$ deviate from those of $\mathrm{WDM}${-}$\mathrm{novth}$ at high wavenumbers, as a consequence of the noisier ICs. As expected from the discussion in Section 2, the effects of the numerical noise depend on the WDM mass. Defining the ratio between the $\mathrm{WDM}${-}$\mathrm{vth}$  and $\mathrm{WDM}${-}$\mathrm{novth}$ matter power spectrum as,
\begin{equation}R(k,z)\equiv \frac{P^{\mathrm{WDM}\mhyphen\mathrm{vth}} (k,z)}{P^{\mathrm{WDM}\mhyphen\mathrm{novth}} (k,z)},
\label{eq:ratiopowerscale}
\end{equation}
$R$ is of the order of $\sim 700$ at redshift $z=99$ and at the Nyquist frequency $k_\mathrm{Ny} \simeq 64$ $h$/Mpc for $m_\mathrm{WDM}=2$ keV. It decreases to $R(k_\mathrm{Ny},99)\sim 30$ for $m_\mathrm{WDM}=3.3$ keV and to $R(k_\mathrm{Ny},99)\sim 1.3$ for $m_\mathrm{WDM}=7$ keV. 

At intermediate redshifts ($z=9-5$), the situation changes. For CDM, the highest wavenumbers probed in our analysis start to enter the nonlinear regime and the ratio of the scaled power spectra starts to deviate from unity. The same effects can be observed in $\mathrm{WDM}${-}$\mathrm{novth}$. Note that since the nonlinear evolution transfers power from large scales to small scales, the differences between CDM and the various WDM models are significantly reduced in the nonlinear matter power spectra \cite{2012MNRAS.421...50V}. Therefore, WDM displays more nonlinear growth than CDM since the WDM initial matter power spectra are highly-suppressed compared to CDM initial $P(k)$ at small scales. At such redshifts, the differences between $\mathrm{WDM}${-}$\mathrm{vth}$  and $\mathrm{WDM}${-}$\mathrm{novth}$ are also reduced. For example, simulations with $m_\mathrm{WDM}=2$ keV show the most dramatic deviations at high redshifts: $R(k_\mathrm{Ny},99)\sim 700$, which become smaller at $z=9$, $R(k_\mathrm{Ny},9)\sim 1.5$, and further reduced to $R(k_\mathrm{Ny},5)\sim 1.07$ at $z=5$. This is because, being random, the noise effect does not significantly grow in size, and when the effect of gravitational collapse starts to grow it gradually dominates over the noise effect. Similar behaviour is found for the other WDM candidates.

At late times ($z\leq 2$), all the wavenumbers probed in our analysis have entered the nonlinear regime, where the nonlinear growth factors for $\mathrm{WDM}${-}$\mathrm{vth}$  and $\mathrm{WDM}${-}$\mathrm{novth}$ are indistinguishable for all three WDM masses. However, small deviations can be seen by looking at the ratio $R$, as shown in the bottom panels of figures \ref{fig:allmassesb}, \ref{fig:allmassesc}, \ref{fig:allmassesd}. As we can see, for all the WDM candidates studied here, the difference in the power spectra of $\mathrm{WDM}${-}$\mathrm{vth}$ with respect to $\mathrm{WDM}${-}$\mathrm{novth}$ is less than $2\%$ below $k_{Ny}$ for redshifts $z\leq2$. For completeness, we mention that a fitting formula exists for the nonlinear matter power spectra for $m_\mathrm{WDM} > 0.5$ keV and redshifts $z<3$ \cite{2012MNRAS.421...50V}. We have checked that our nonlinear power spectra agree with this parametrisation with the $2\%$ accuracy claimed in \cite{2012MNRAS.421...50V}.

\begin{figure}[]
\advance\leftskip-1.cm
\advance\rightskip-2cm
\subfigure[][CDM.]
{\includegraphics[width=.62\textwidth,height=7.5cm]{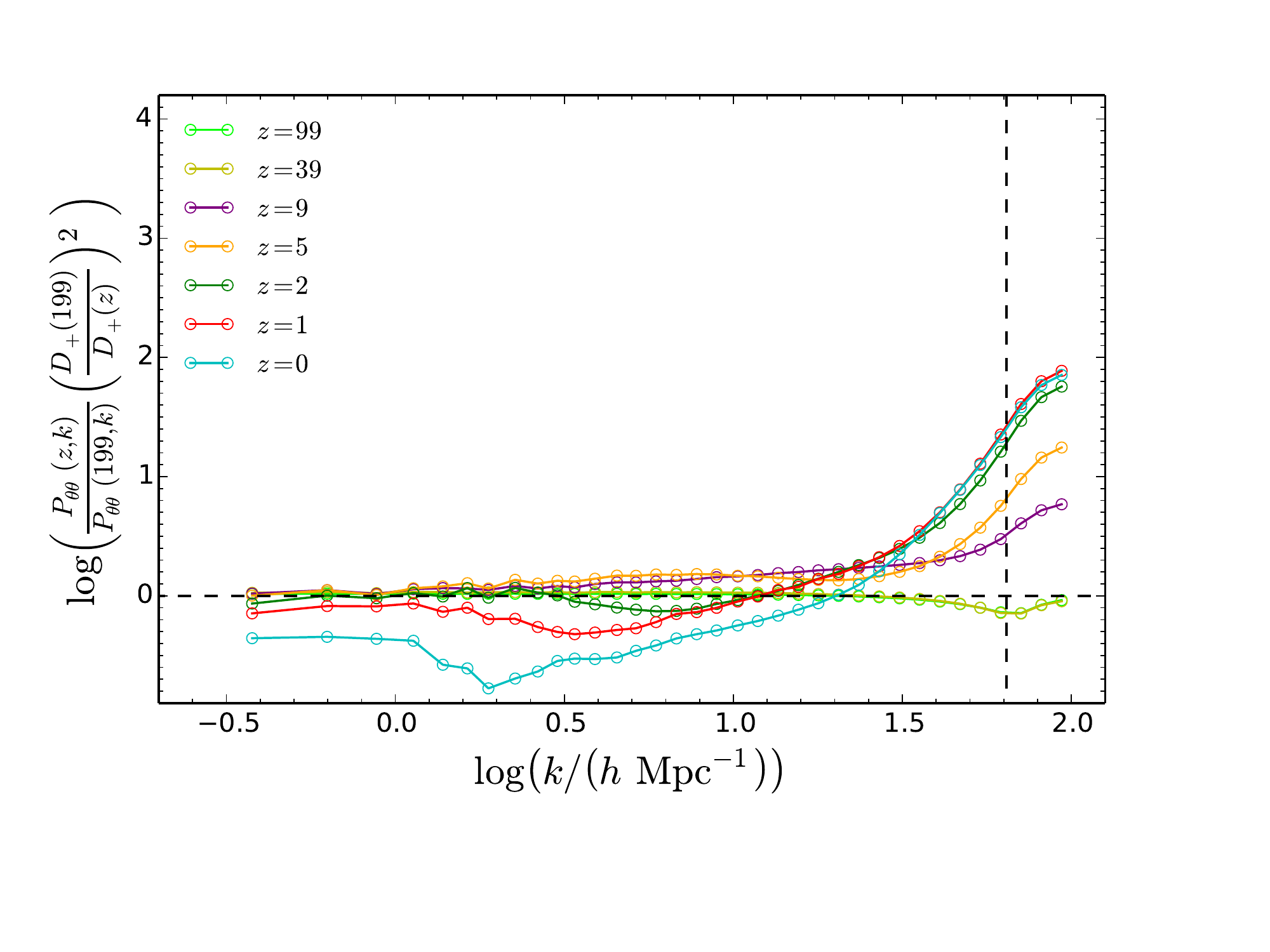}\label{fig:allvelmassesa}}  \hspace{-3.1\baselineskip}
\subfigure[][$m_\mathrm{WDM} = 7$ keV.]
{\includegraphics[width=.62\textwidth,height=7.5cm]{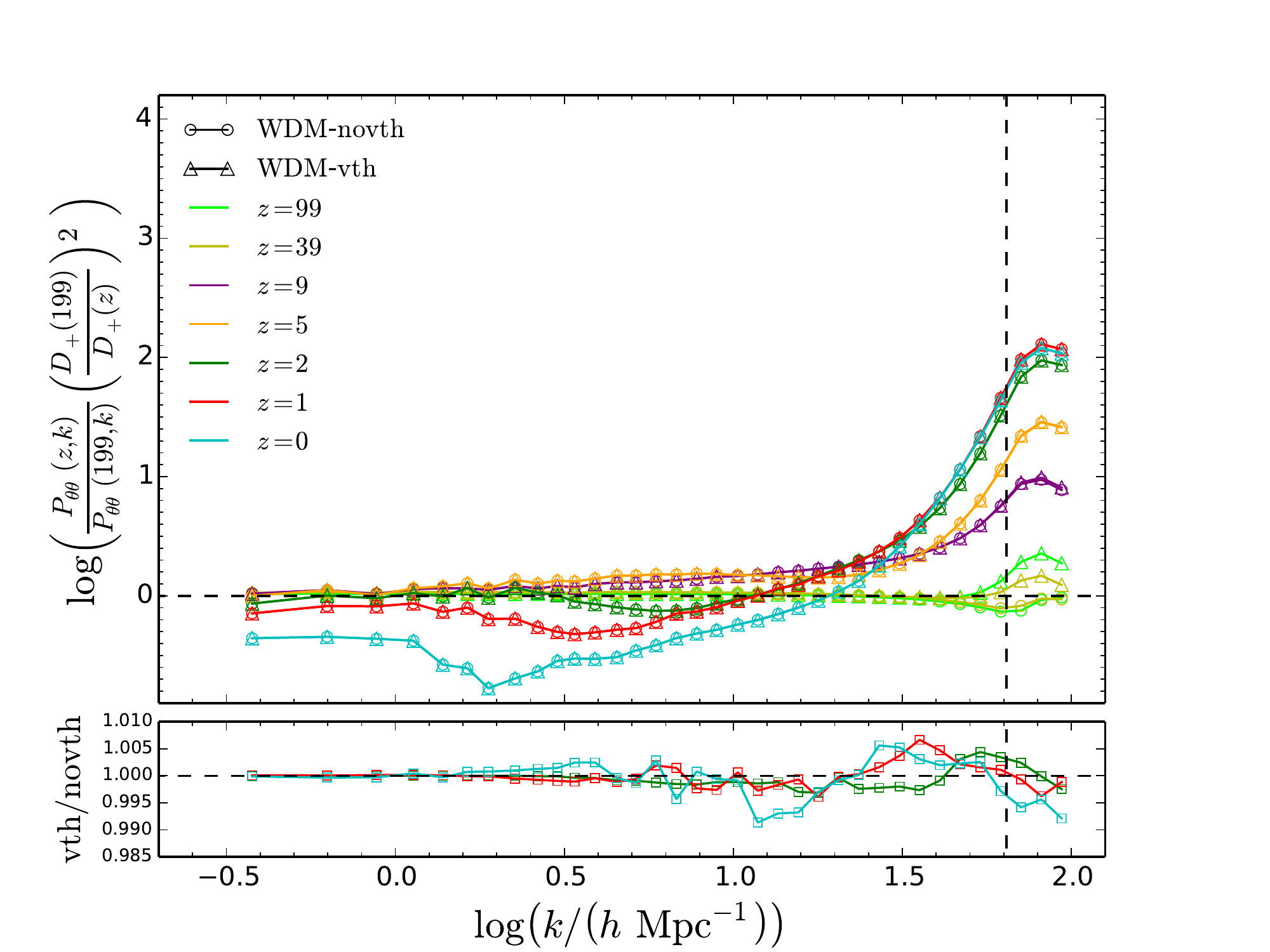}\label{fig:allvelmassesb}} \\
\subfigure[][$m_\mathrm{WDM} = 3.3$ keV.]
{\includegraphics[width=.62\textwidth,height=7.5cm]{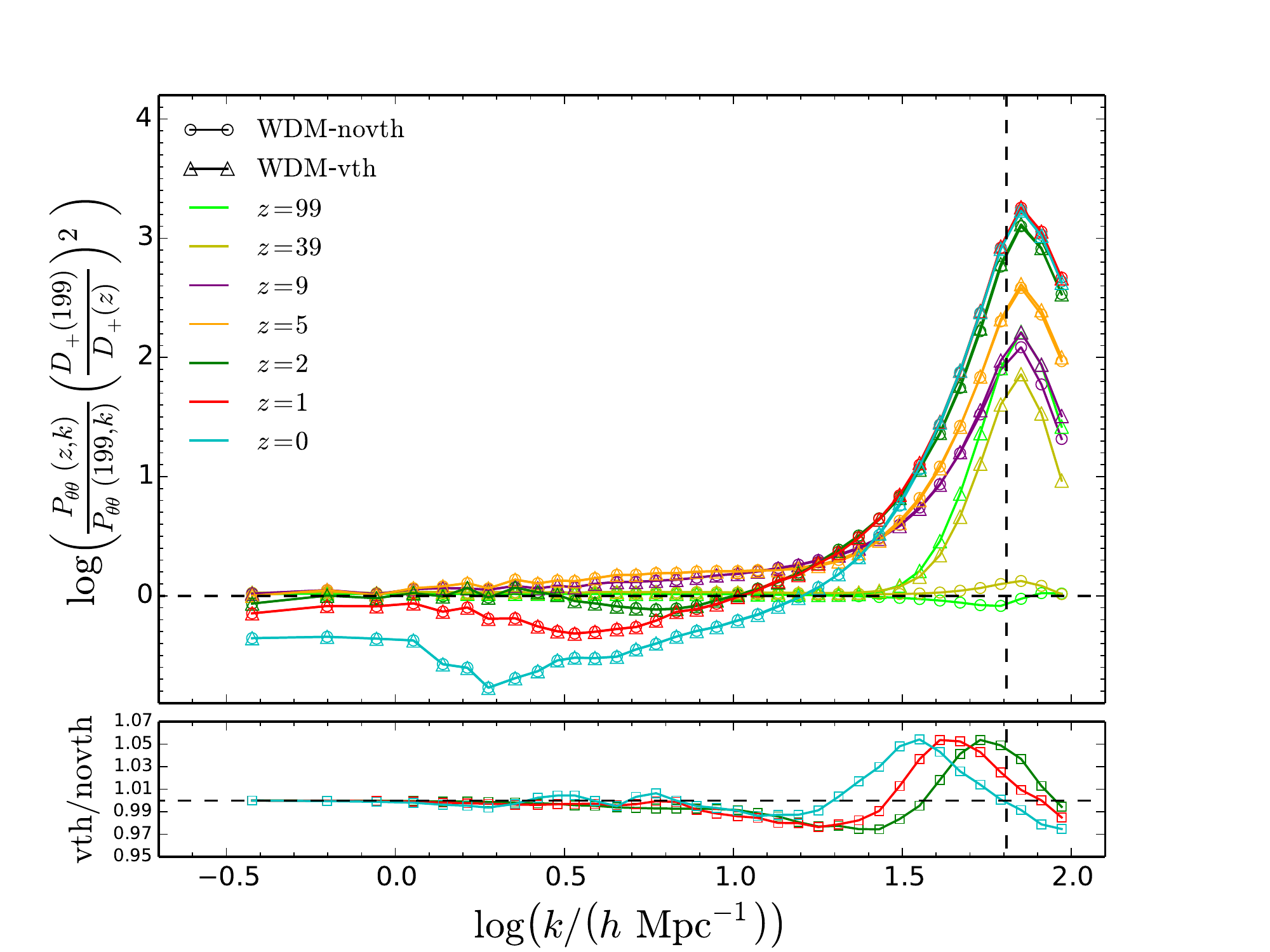}\label{fig:allvelmassesc}}  \hspace{-3.1\baselineskip}
\subfigure[][$m_\mathrm{WDM} = 2$ keV.]
{\includegraphics[width=.62\textwidth,height=7.5cm]{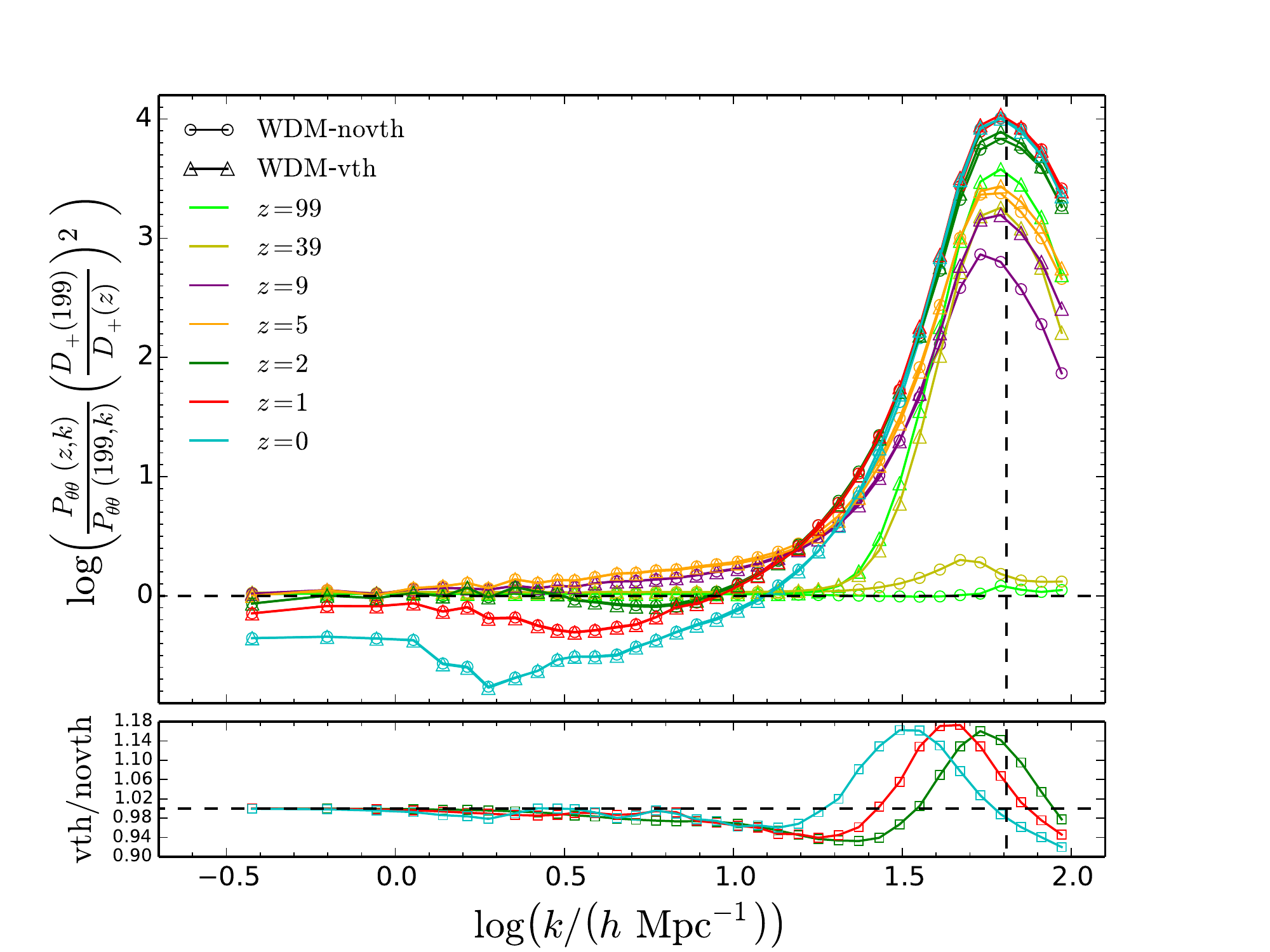}\label{fig:allvelmassesd}} \\   
\caption{The same as Figure \ref{fig:allmasses}, but for velocity power spectra. All spectra are divided by the initial noiseless spectrum at $z=199$ (the velocity spectrum measured from the ICs for $\mathrm{WDM}${-}$\mathrm{novth}$).}
\label{fig:allvelmasses}
\end{figure}

We have performed a similar analysis for the velocity power spectra. Since the thermal velocity noise also affects the initial redshift $z_\mathrm{ini}=199$, in our plots we choose to take the nonlinear growth $\left(P(z)/P(199)\right) \left(D_+(199)/D_+(z)\right)^2$ by dividing all the power spectra ($\mathrm{WDM}${-}$\mathrm{vth}$  and $\mathrm{WDM}${-}$\mathrm{novth}$) by the initial noiseless velocity power spectrum $P(z_\mathrm{ini}=199)$. Figure  \ref{fig:allvelmasses} shows the results for CDM and the three WDM simulations. The noise effect in $\mathrm{WDM}${-}$\mathrm{vth}$ is dominant at high redshifts. For example, the ratio $R$ for the velocity power spectrum is around $R(k_\mathrm{Ny},99) \sim 3000$ for $m_\mathrm{WDM} = 2$ keV. The impact of the noise depends on the mass and $R$ is smaller for the two colder WDM candidates: $R(k_\mathrm{Ny},99) \sim 100$ for $m_\mathrm{WDM} = 3.3$ keV and $R(k_\mathrm{Ny},99) \sim 1.8$ for $m_\mathrm{WDM} = 7$ keV. 

At intermediate ($z=9-5$) and low redshifts ($z\leq2$), the highest wavenumbers probed in our simulations have entered the non-linear regime. However, the transfer of power from large to small scales is more pronounced for the velocity $P(k)$, with the power transfer wavenumber moving to higher $k$. At these redshifts the effects due to the noise in the ICs of $\mathrm{WDM}${-}$\mathrm{vth}$ are strongly reduced. For the warmest candidate  ($m_\mathrm{WDM} = 2$ keV) we find that $R(k_\mathrm{Ny},9)\sim 2.5$ at $z=9$, which is further reduced to $R(k_\mathrm{Ny},5)\sim1.2$ at $z=5$. Similar behaviour is found for the other WDM candidates.

The values of the ratio $R$ at the lowest redshifts $z=2,1,0$ are plotted in the bottom panels of figures \ref{fig:allvelmassesb}, \ref{fig:allvelmassesc}, \ref{fig:allvelmassesd}. As we can see, unlike the matter power spectra, in the velocity power spectra $R$ does not decrease at $z\leq2$, but rather its peak shifts to lower $k$ when $z$ decreases. For example, for the warmest candidate, $R$ has a peak at $k\sim 50$ $h/$Mpc at $z=2$ and it is shifted to small wavenumbers, $k\sim 32$ $h/$Mpc, at $z=0$, while the magnitude of the peak does not change and is around $18\%$ for all $z\leq 2$. For $m_\mathrm{WDM} =3.3$ keV, we find the same result, but in this case the magnitude of the peak is around $R\sim6 \%$ for all $z\leq2$. For $m_\mathrm{WDM} =7$ keV, $R$ is always well below $1 \%$ for redshifts $z\leq2$. 

In conclusion, the matter and velocity power spectra for $\mathrm{WDM}${-}$\mathrm{vth}$ are always affected by the noise. The importance of the noise is reduced when the system evolves and the nonlinear gravitational evolution starts to dominate. However, we note that in the warmest WDM scenario examined here ($m_\mathrm{WDM}=2$ keV), the noise still affects appreciably the velocity power spectra at $z\leq2$, causing deviations of the order of $\sim 18\%$. Therefore, adding thermal velocities reduces the range of validity of the simulation predictions. We mention here that we have also performed a resolution analysis (similar to what we have done in Section 2 and which we omit for brevity) on the evolved matter and velocity power spectra, and we have checked that the differences in the evolved power spectra for simulations with thermal velocities are truly due to numerical noise.

\subsection{Results for initial conditions generated at $z_\mathrm{ini}=39$}

It was suggested by \cite{Colin:2007bk} for WDM candidates and \cite{Brandbyge:2008rv,Viel:2010bn} for active neutrinos that a useful way to suppress the effect of the thermal velocity noise in the matter power spectrum is to start the simulation at a lower redshift, when the thermal velocity contribution is smaller (see eq. (\ref{eq:rmsthermalvel})). To check this, we have run another set of simulations with initial redshift $z_\mathrm{ini}=39$. We have evolved the two set of ICs (the one at $z_\mathrm{ini}=199$ and the other at $z_\mathrm{ini}=39$) to $z=19$ and measured the matter power spectra from the snapshots at $z=39$ and $z=19$.

\begin{figure}
\centering
\subfigure[][ICs at $z=199$.]
{\includegraphics[width=.52\textwidth]{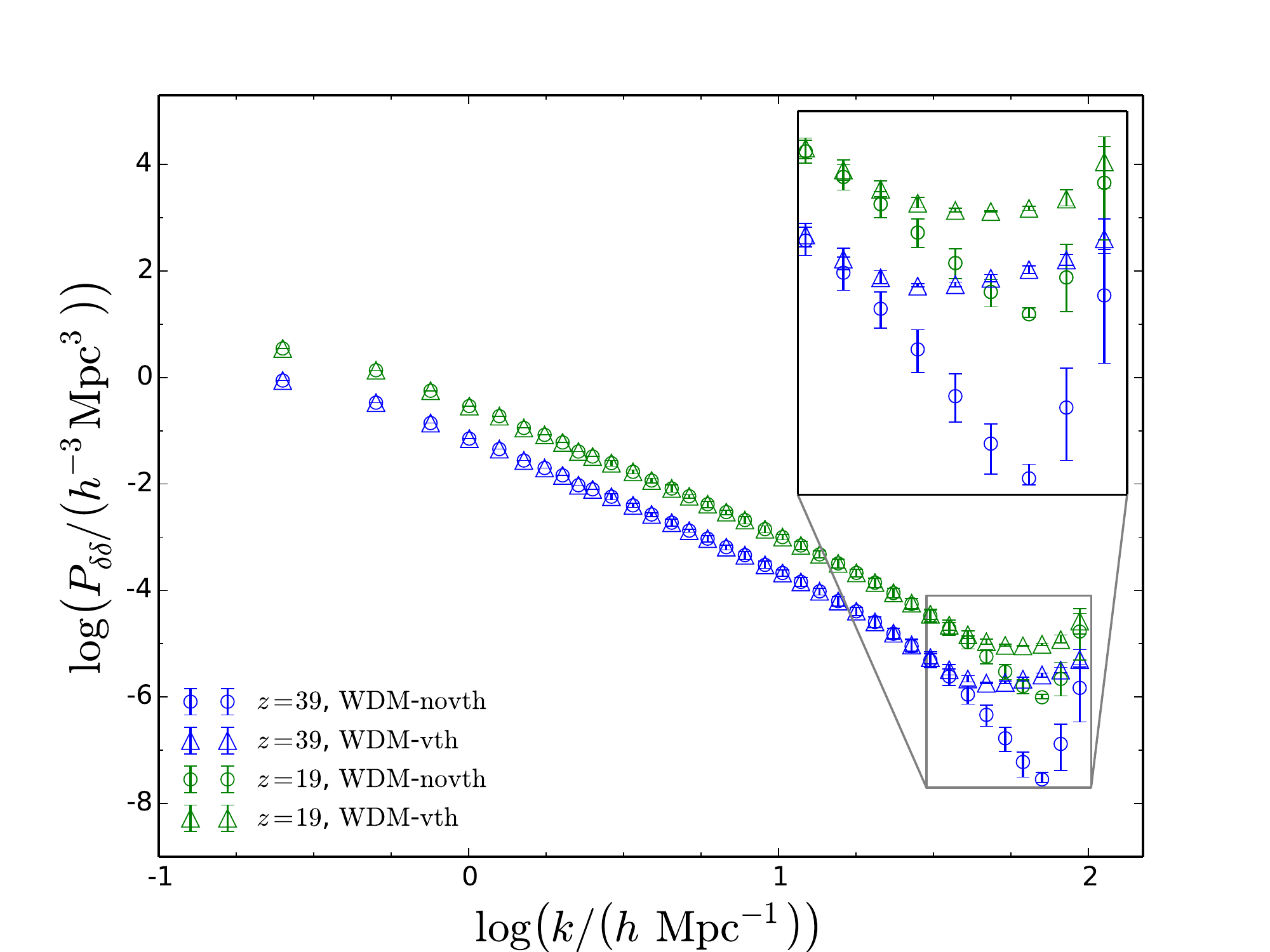}\label{fig:z39suma}} \hspace{-2.2\baselineskip}
\subfigure[][ICs at $z=39$.]
{\includegraphics[width=.52\textwidth]{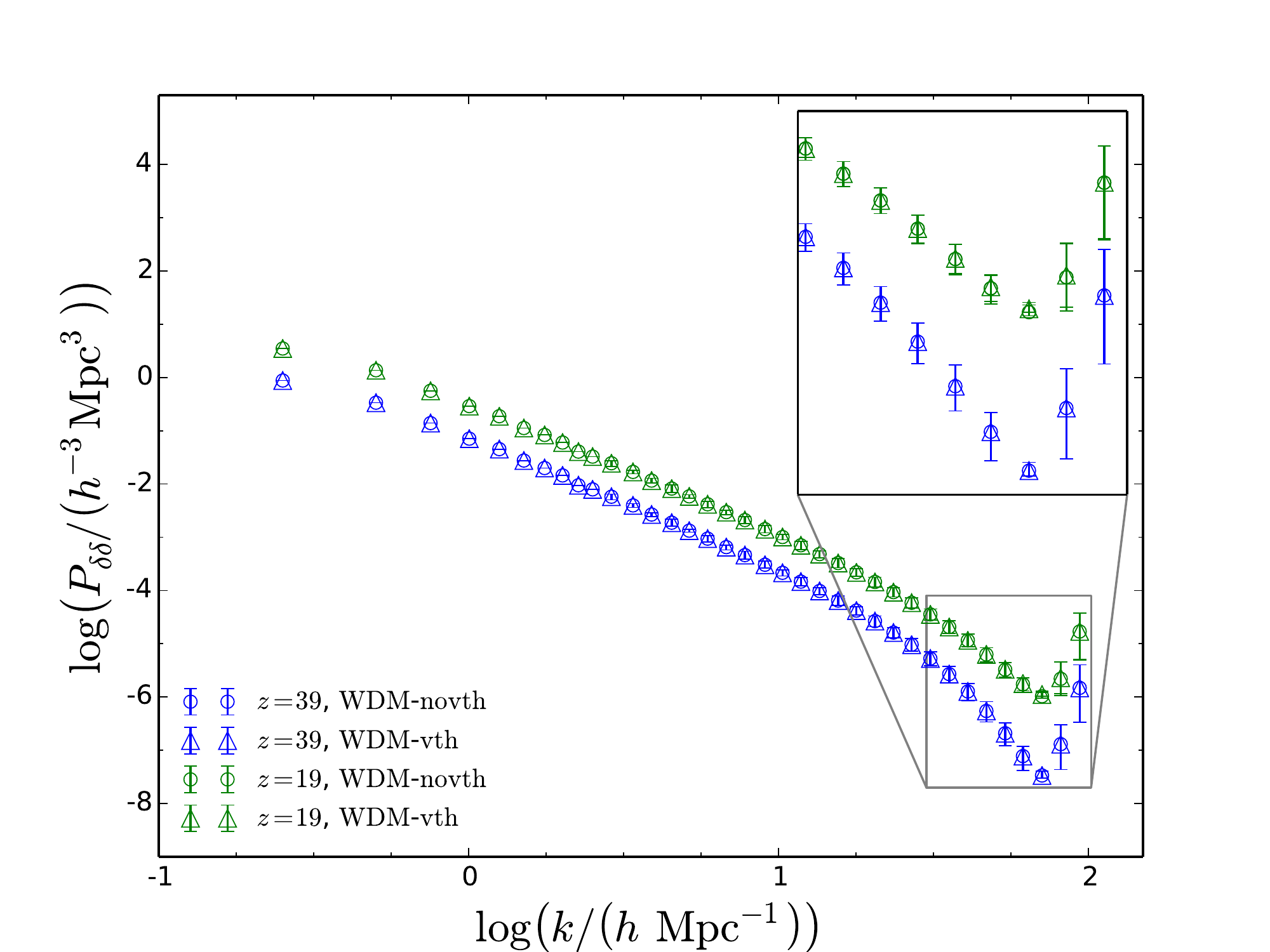}\label{fig:z39sumb}} \\
\caption{(a) Matter power spectra at redshifts $z=39, 19$ for WDM with $m_w = 3.3$ keV with the ICs at $z_\mathrm{ini}=199$. (b) Matter power spectra at redshifts $z=39, 19$ for WDM with $m_w = 3.3$ keV with the ICs at $z_\mathrm{ini}=39$.}
\label{fig:z39sum}
\end{figure}

The results for $m_\mathrm{WDM} = 3.3$ keV are shown in figure \ref{fig:z39sum}. The ratio $R$ given in eq. (\ref{eq:ratiopowerscale}) is of the order of $\sim 30$ at $z=39$ and at the Nyquist frequency $k_\mathrm{Ny}\sim 64$ $h/$Mpc for the first set of simulations with starting redshift at $z_\mathrm{ini}=199$, while it is exactly 1 for all the wavenumbers probed in the simulations with the ICs at $z_\mathrm{ini}=39$. This is because the $z=39$ snapshots are the ICs for the second set of simulations and, as mentioned in Section 2, the matter power spectra measured from the ICs are not affected by the noise. On the other hand, the (evolved) matter power spectra at $z=19$ are affected by the noise in both the two set of $\mathrm{WDM}${-}$\mathrm{vth}$ simulations, but the choice of a lower starting redshift reduces the differences with respect to $\mathrm{WDM}${-}$\mathrm{novth}$. Indeed, the matter power spectrum measured from $\mathrm{WDM}${-}$\mathrm{vth}$ with the ICs at $z_\mathrm{ini}=199$ is affected by the noise and deviates significantly  at high wavenumbers from the result measured from $\mathrm{WDM}${-}$\mathrm{novth}$  at $z=19$. Quantitatively, we have $R\sim 10$ at $z=19$ and near $k_\mathrm{Ny}$ for simulations with the ICs at $z_\mathrm{ini}=199$. However, in the case of ICs at $z_\mathrm{ini}=39$ the differences in the matter power spectra when adding thermal velocities are negligible, e.g. at $z=19$ we have $R \lesssim 1.04$ for $k < k_\mathrm{Ny}$. The same behaviour is found at lower redshifts and for velocity power spectra, which are not presented here. This test confirms that the enhanced power observed in simulations with thermal velocities is numerical and can be reduced by choosing a lower initial redshift. 

By starting at a lower redshift we inevitably lose information about the high redshift evolution of the system and the accuracy of the simulations could be compromised. This is because we need to evolve the simulation a large number of expansion factors away from ICs to get the proper gravitational evolution of the density field. However, using second-order Lagrangian perturbation theory (on which the code 2LPTic is based) the accuracy of the simulations without thermal velocities starting at low redshift ($z_\mathrm{ini}=39$) is not seriously compromised for the scales resolved in our simulations. For example, considering the two power spectra at $z=19$ displayed in figure \ref{fig:z39sum} (those measured from $\mathrm{WDM}${-}$\mathrm{novth}$ with ICs at $z_\mathrm{ini}=199$ and from $\mathrm{WDM}${-}$\mathrm{novth}$ with ICs at $z_\mathrm{ini}=39$ respectively), we find that they agree between each other up to $k\sim 10\,h/\mathrm{Mpc}$. Above such wavenumbers the $P(k)$ with ICs at $z_\mathrm{ini}=39$ displays more power than that with ICs at $z_\mathrm{ini}=199$, although the differences in the amplitudes between the two power spectra never exceed $\sim 10 \%$. So, starting at low redshift does not affect the results significantly from simulations without thermal velocities, while it drastically reduces the extra numerical noise introduced when adding thermal velocities.

\section{Halo properties}
As found in the previous section, the numerical noise caused by including thermal velocities influences even the lowest redshifts. We expect that some halo properties can also be affected by this noise. In this section, we quantify how the $\mathrm{WDM}${-}$\mathrm{vth}$ simulations differ from $\mathrm{WDM}${-}$\mathrm{novth}$ in halo properties, such as the halo mass function and halo density profiles.
\subsection{Halo mass functions}
It is well known that a reduction in power on small scales is reflected in the suppression of the number of low mass structures in the Universe (once spurious structures have been  removed) \cite{Bode:2000gq,Wang:2007he,Lovell:2013ola,Schewtschenko:2014fca,2012MNRAS.424..684S,Power:2013rpw,Power:2016usj}. This can be seen by counting the number of haloes as a function of mass (i.e. the halo mass function) in WDM simulations, and comparing with the result from a CDM simulation. As we saw in Section 3, the noise in the ICs due to thermal velocities propagates to late times and could influence the halo mass function. To study this we focus on simulation snapshots at $z=0$. To extract the halo properties, we use the code {\sc rockstar}, which is a phase-space friend-of-friends halo finder \cite{2013ApJ...762..109B}. We use the virial mass ($M_\mathrm{vir}$) to characterise the haloes, which is defined as the mass enclosed in a sphere of radius $r_\mathrm{vir}$, where $r_\mathrm{vir}$ is the virial radius given in \cite{Bryan:1997dn}. The (differential) halo mass function is presented as $F(M_\mathrm{vir}, z) = dn/d\log(M_\mathrm{vir})$, where $n$ is the number density of haloes with virial mass $M_\mathrm{vir}$. 

As mentioned in Section 3, WDM simulations (in which the initial power spectrum has a resolved cut-off) are affected by the artificial fragmentation of filaments, i.e. spurious haloes, which can be seen as an upturn in the halo mass function at small masses. A mass cut-off, $M_{\text{lim}}$, was proposed in \cite{Wang:2007he} below which haloes are likely to be spurious\footnote{This limit can be considered as an estimate. Not all of the haloes with masses below this limit are unphysical. Furthermore there could be some spurious haloes with masses above this limit. The common way to find the unphysical haloes is explained in \cite{Lovell:2013ola}. Since our treatment is simply intended to show the effects of adding thermal velocities, we will not go into a detailed study of how to eliminate spurious haloes.}:
\begin{equation}
M_{\text{lim}} = 10.1 \, \bar{\rho}\, d\, k_{\text{peak}}^{-2},
\label{eq:WandW}
\end{equation}
where $\bar{\rho}$ is the mean density of the Universe, $d$ is the mean interparticle separation and $k_{\text{peak}}$ is the wavenumber at which the dimensionless matter power spectrum, $\Delta(k) \equiv k^3 P(k)/(2\pi^2)$, has its maximum. The simulations used by \cite{Wang:2007he} did not include thermal velocities.

\begin{figure}
\centering
\subfigure[][$m_\mathrm{WDM} =  2$ keV, $L = 25$ $h^{-1}$Mpc]
   {\includegraphics[width=.48\textwidth]{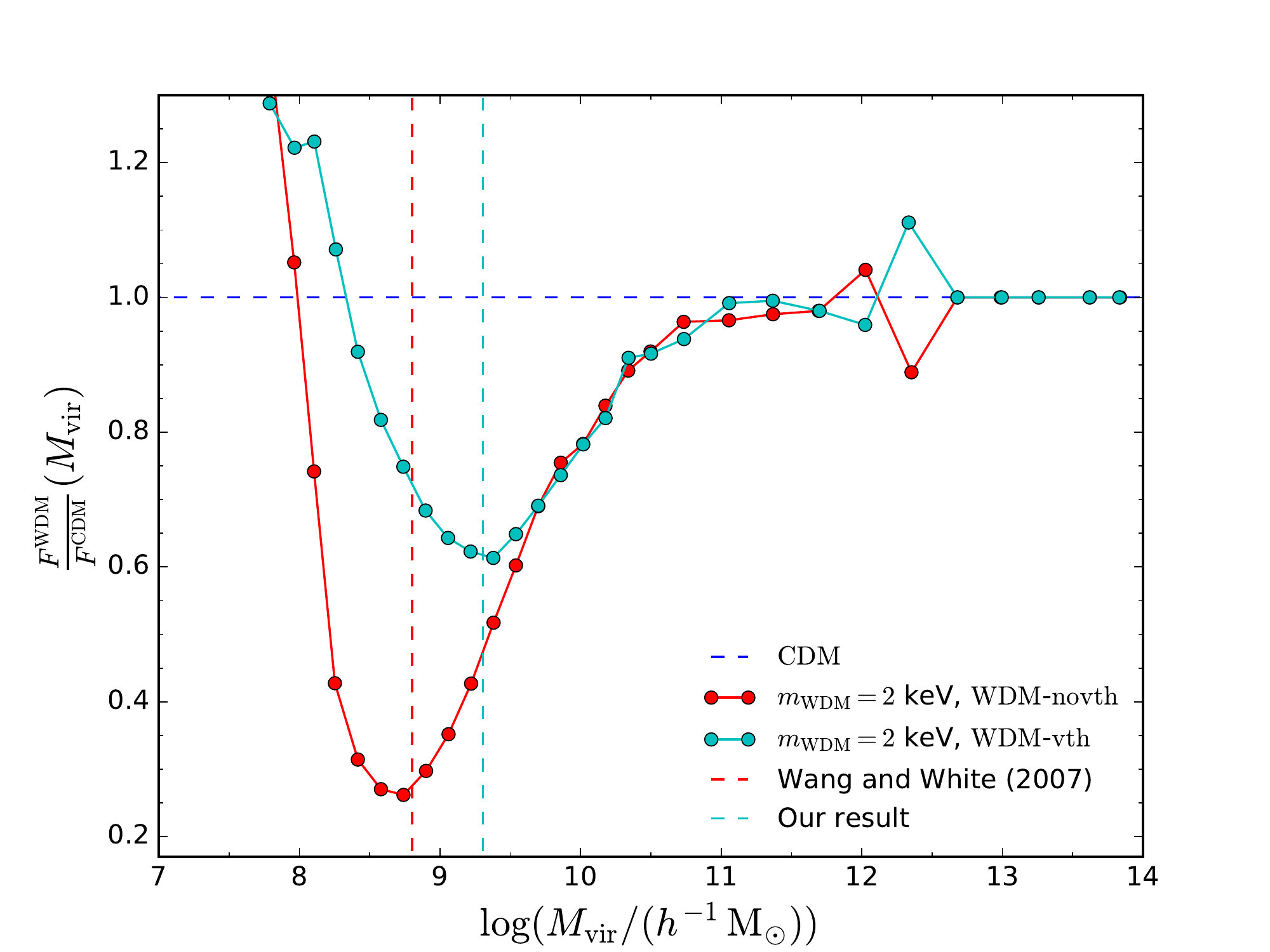}\label{fig:halosa}} \quad
\subfigure[][$m_\mathrm{WDM} =  3.3$ keV, $L = 25$ $h^{-1}$Mpc]
   {\includegraphics[width=.48\textwidth]{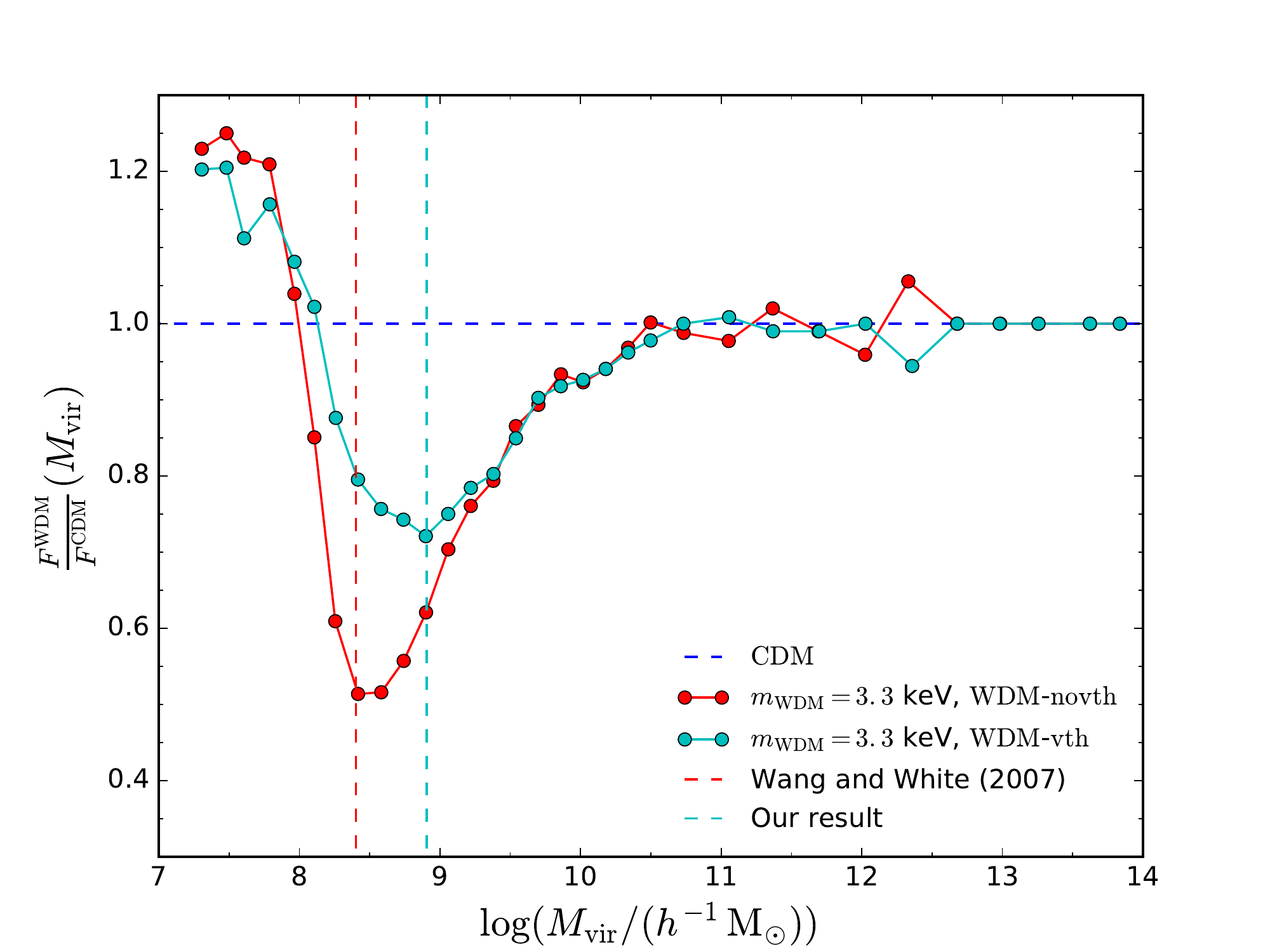}\label{fig:halosb}} \quad
\subfigure[][$m_\mathrm{WDM} =  7$ keV, $L = 25$ $h^{-1}$Mpc]
   {\includegraphics[width=.48\textwidth]{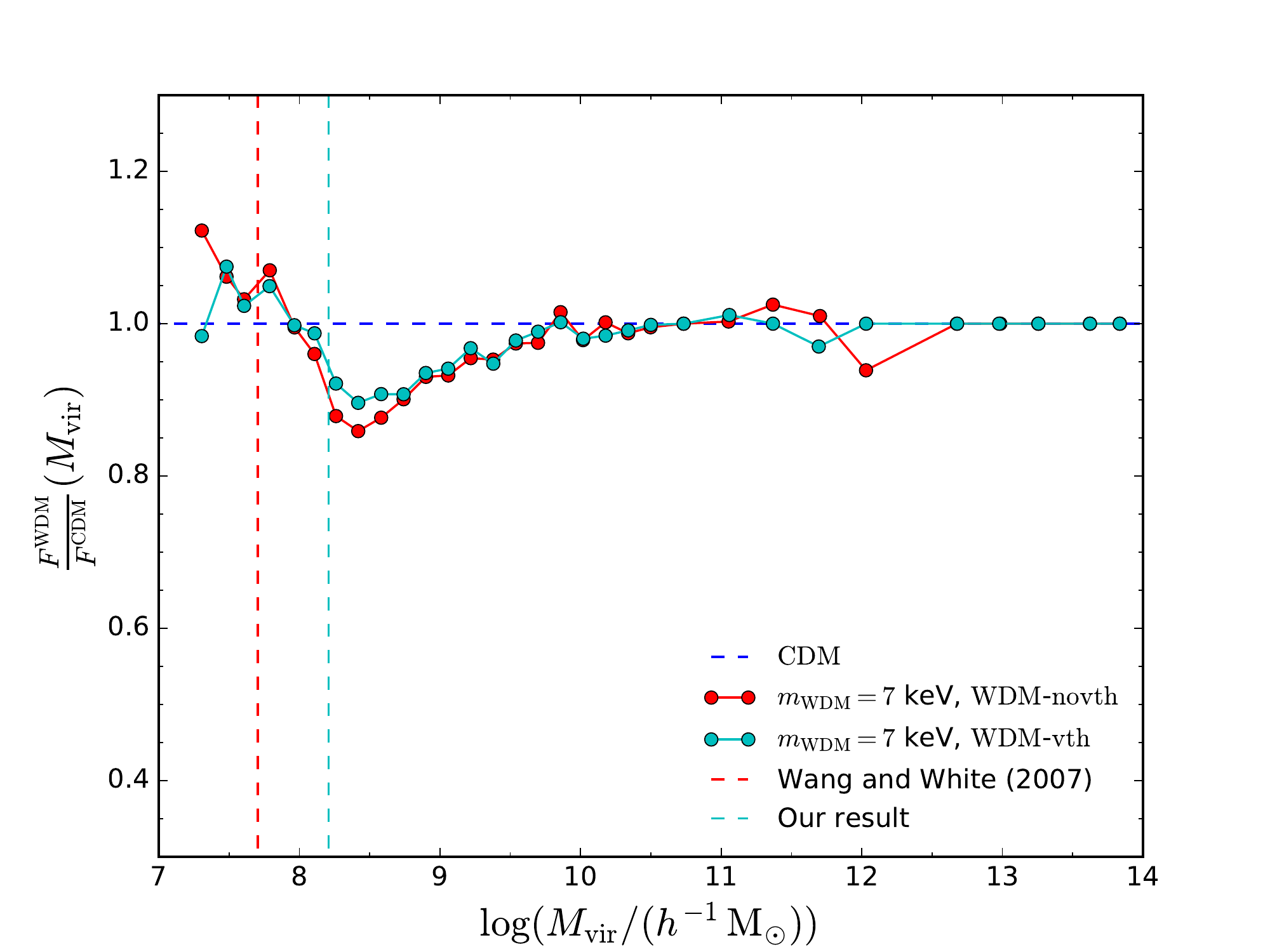}\label{fig:halosc}} \quad     
\subfigure[][$m_\mathrm{WDM} =  7$ keV, $L = 12$ $h^{-1}$Mpc]
   {\includegraphics[width=.48\textwidth]{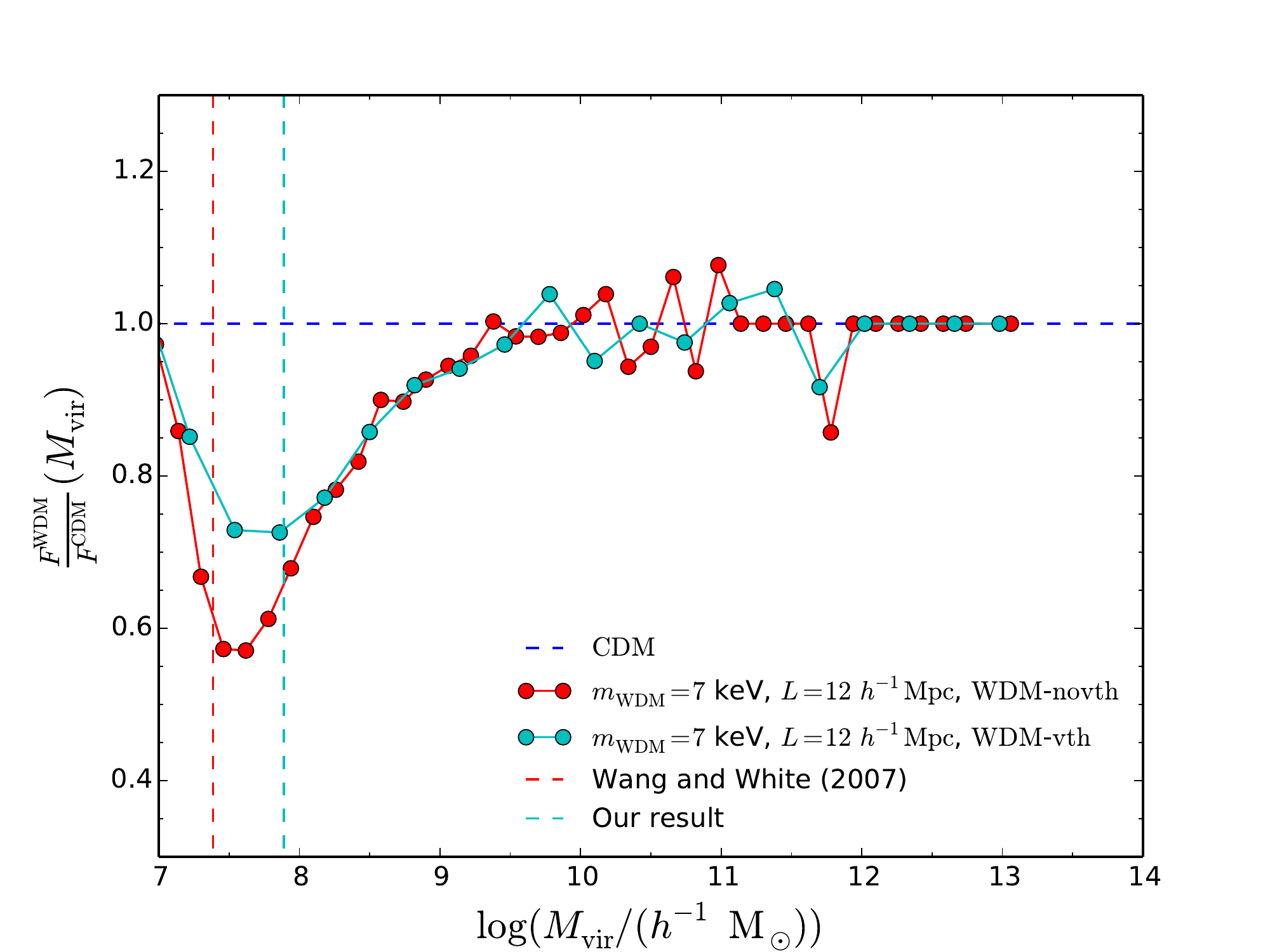}\label{fig:halosd}} \quad     
\caption{Halo mass function for WDM plotted as a ratio with respect to CDM. $\mathrm{WDM}${-}$\mathrm{vth}$ are in cyan, while $\mathrm{WDM}${-}$\mathrm{novth}$ are in red. The red vertical dashed line is the mass cut-off suggested in \cite{Wang:2007he} (see eq. (\ref{eq:WandW})) to take care of the spurious haloes. The cyan vertical dashed line is our limit for simulations with thermal velocities (see eq. (\ref{eq:WandWmod})). The halo mass functions are measured from the simulations at redshift $z=0$. For (a) (b) (c) the simulation box size is $L = 25$ $h^{-1}$Mpc. For the $m_\mathrm{WDM} = 7$ keV candidate we ran an extra simulation (d) with a smaller box size,  $L = 12$ $h^{-1}$Mpc.}
\label{fig:8}
\label{fig:halos}
\end{figure}

In figure \ref{fig:halos} we plot the ratio between the halo mass functions measured from our WDM and CDM simulations. As we can see in the figure, at the largest masses probed by our simulations the WDM and CDM models predict the same number of haloes. However, the halo abundance in WDM is progressively reduced at smaller masses. The mass scales at which the damping of the structures appears depend on the WDM particle mass. Colder candidates shift the damping in the halo mass function towards smaller masses. However, all the WDM halo mass functions in figure \ref{fig:halos} show a clear upturn at low masses, which is evidence of spurious haloes. For $\mathrm{WDM}${-}$\mathrm{novth}$, the mass scale at which the upturn appears is well described by eq. (\ref{eq:WandW}) (see the red vertical dashed line in figures \ref{fig:halosa} and \ref{fig:halosb}; the results for the candidate with $m_\mathrm{WDM}=7$ keV will be treated separately). However, eq. (\ref{eq:WandW})  fails to reproduce the position of the upturn seen in our $\mathrm{WDM}${-}$\mathrm{vth}$ simulations. As we can see, there are more (spurious) substructures at low masses in $\mathrm{WDM}${-}$\mathrm{vth}$ and these spurious haloes appear at larger masses than in $\mathrm{WDM}${-}$\mathrm{novth}$.  These differences are caused by the additional noise present in $\mathrm{WDM}${-}$\mathrm{vth}$. We have found a revised mass cut-off for $\mathrm{WDM}${-}$\mathrm{vth}$ (see the cyan vertical dashed line in figure \ref{fig:halos}),  
\begin{equation}
M_{\text{lim}} = 32.2\, \bar{\rho}\, d\, k_{\text{peak}}^{-2}.
\label{eq:WandWmod}
\end{equation}

Eq.~(\ref{eq:WandW}) (for $\mathrm{WDM}${-}$\mathrm{novth}$) and our result (\ref{eq:WandWmod}) (for $\mathrm{WDM}${-}$\mathrm{vth}$) reproduce quite well the cut-off scales for WDM candidates with masses $m_\mathrm{WDM} =2$ keV and $3.3$ keV. However, both formulae fail to reproduce the predicted mass cut-off for the coldest candidate ($m_\mathrm{WDM} =7$ keV) in a simulation with a box length of $25$ $h^{-1}$Mpc and $512^3$ particles (see figure \ref{fig:halosc}). The reason is that this simulation has a Nyquist frequency $k_\mathrm{Ny} \sim 64$ $h/$Mpc, which is below the half-mode wavenumber $k_\mathrm{hm} \sim 83$ $h/$Mpc (see figure \ref{fig:halfmodeks}) for the case of $m_\mathrm{WDM} =7$ keV. The simulation does not fully resolve the power spectrum cut-off for such heavy WDM masses, and consequently only produces deviations from CDM at very small mass scales, as shown in figure \ref{fig:halosc}. To study the effects on the halo mass function for $m_\mathrm{WDM} =7$ keV  we ran an additional simulation with a box length of $12$ $h^{-1}$Mpc and $512^3$ particles, which has a Nyquist frequency $k_\mathrm{Ny} \sim 134$ $h/$Mpc. The results are shown in figure \ref{fig:halosd}, where we can see that the two formulae (\ref{eq:WandW}) and (\ref{eq:WandWmod}) describe the mass cut-off in the high simulation run very well.

Although the results are presented for $z=0$, we have found that the mass below which spurious haloes become important is the same at higher redshifts, i.e. the spurious haloes appear at roughly the same mass independently of the redshift considered. This is true for simulations with and without thermal velocities.

\subsection{Radial density profiles}
In addition to the halo mass function, we investigate if the radial density profile of haloes in WDM simulations is affected by numerical noise. It is well known that CDM haloes exhibit cuspy profiles \cite{Dubinski:1991bm,Navarro:1995iw,Navarro:1996gj}, described by the NFW fitting formula \cite{Navarro:1996gj},

\begin{equation}
\frac{\rho_\mathrm{fit}(r)}{\rho_c} = \frac{\delta_c}{\frac{r}{r_s}  \left(1+\frac{r}{r_s}\right)^2 },
\label{eq:NFW}
\end{equation}
where $\delta_c$ is the characteristic overdensity and $r_s$ is the scale radius.

\begin{figure}
\centering
\subfigure[][$\log_{10}(M_{halo}/ h^{-1}\text{M}_{\odot}) = (11.8 \pm 0.2)$, averaged over 101 haloes for CDM, 102 for $\mathrm{WDM}${-}$\mathrm{novth}$ and 100 for $\mathrm{WDM}${-}$\mathrm{vth}$.]
   {\includegraphics[width=.48\textwidth]{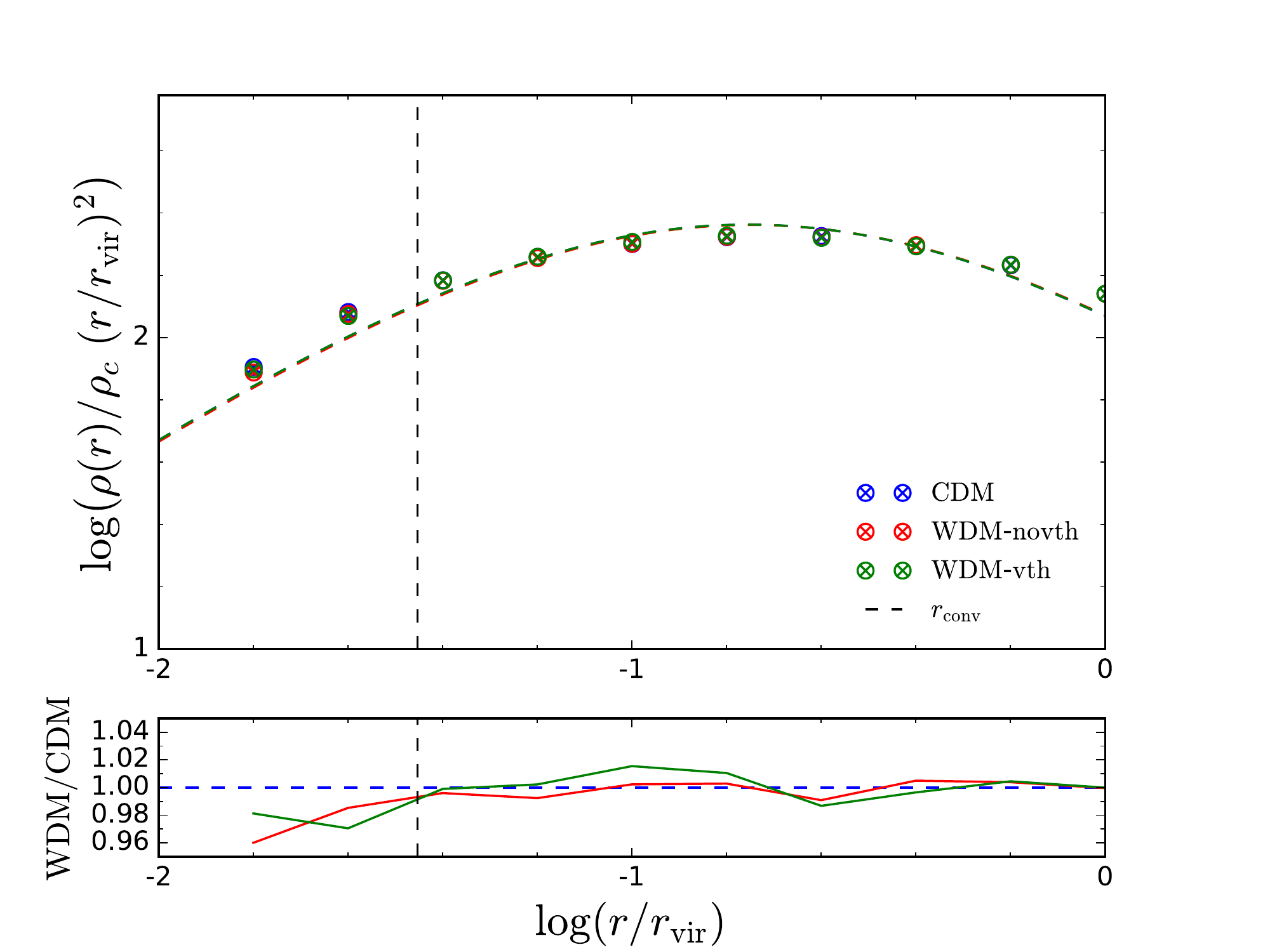}\label{fig:1a33haloesok}} \quad
\subfigure[][$\log_{10}(M_{halo}/ h^{-1}\text{M}_{\odot}) = (10 \pm 0.2)$, averaged over 4104 haloes for CDM, 3867 for $\mathrm{WDM}${-}$\mathrm{novth}$ and 3862 for $\mathrm{WDM}${-}$\mathrm{vth}$.]
   {\includegraphics[width=.48\textwidth]{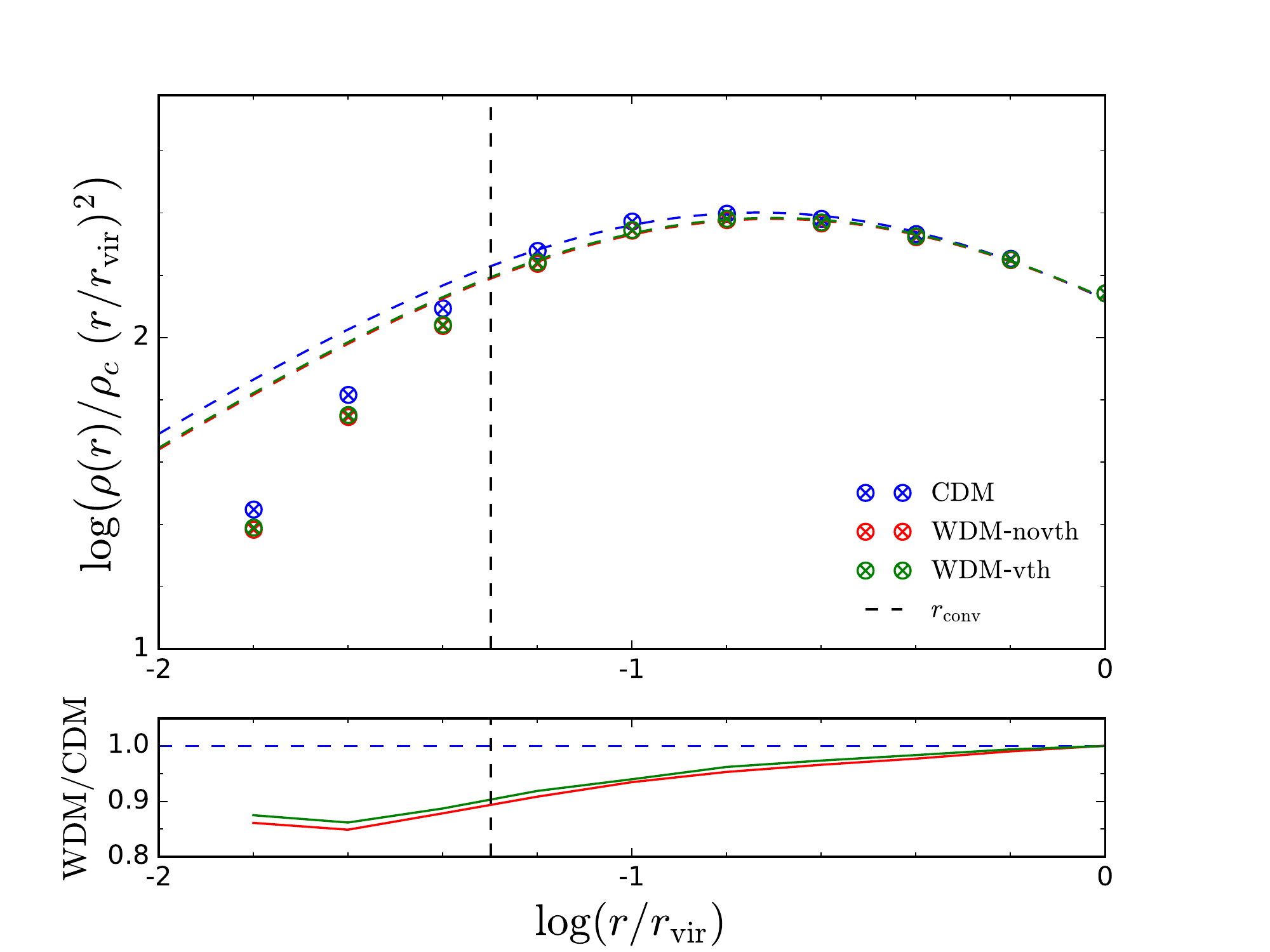}\label{fig:1c33haloesok}} \\   
\subfigure[][$\log_{10}(M_{halo}/ h^{-1}\text{M}_{\odot}) = (11.8 \pm 0.2)$, averaged over 101 haloes for CDM, 103 for $\mathrm{WDM}${-}$\mathrm{novth}$ and 103 for $\mathrm{WDM}${-}$\mathrm{vth}$.]
   {\includegraphics[width=.48\textwidth]{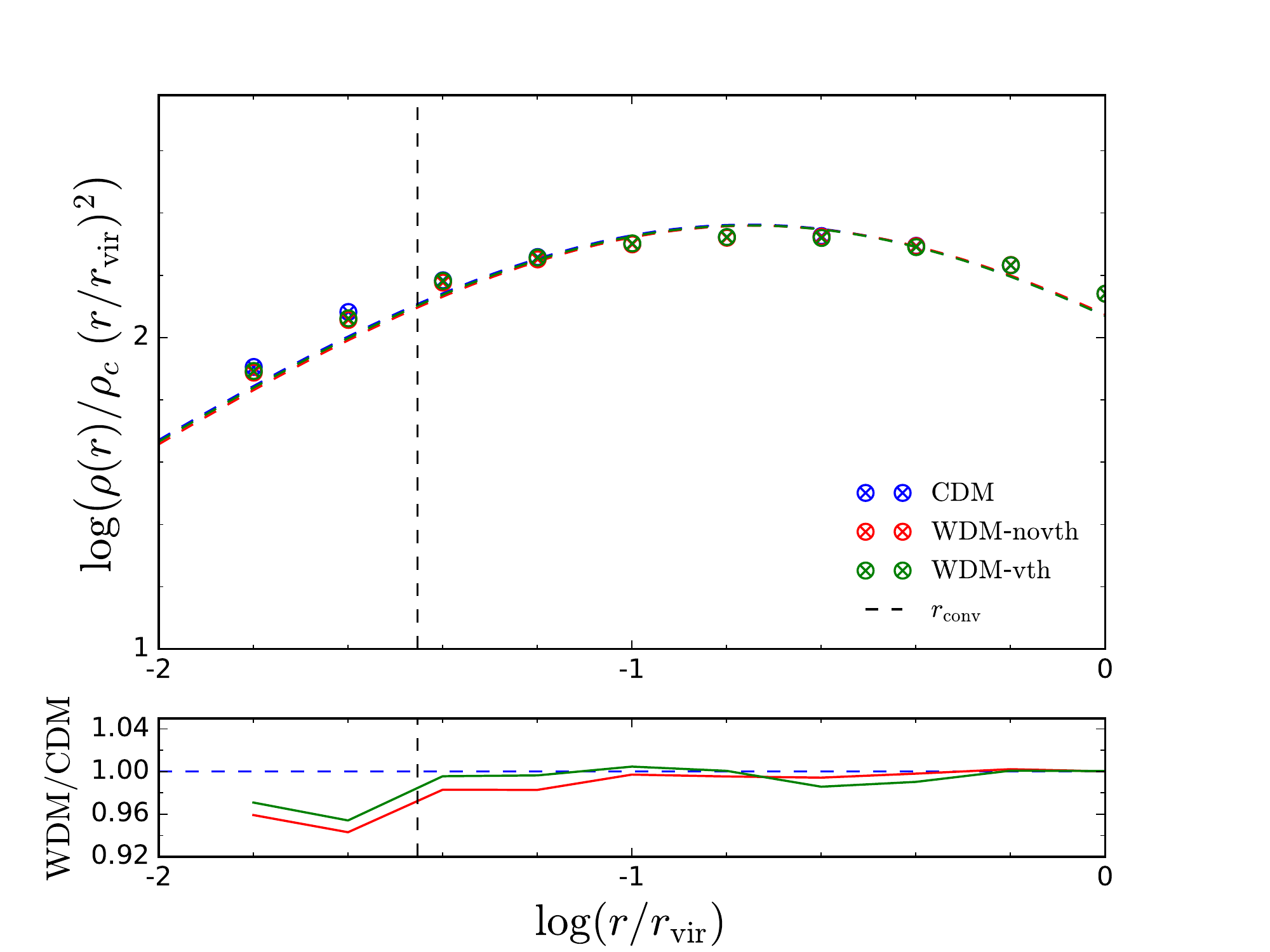}\label{fig:1a2haloesok}} \quad
\subfigure[][$\log_{10}(M_{halo}/ h^{-1}\text{M}_{\odot}) = (10 \pm 0.2)$, averaged over 4104 haloes for CDM, 3256 for $\mathrm{WDM}${-}$\mathrm{novth}$ and 3201 for $\mathrm{WDM}${-}$\mathrm{vth}$.]
   {\includegraphics[width=.48\textwidth]{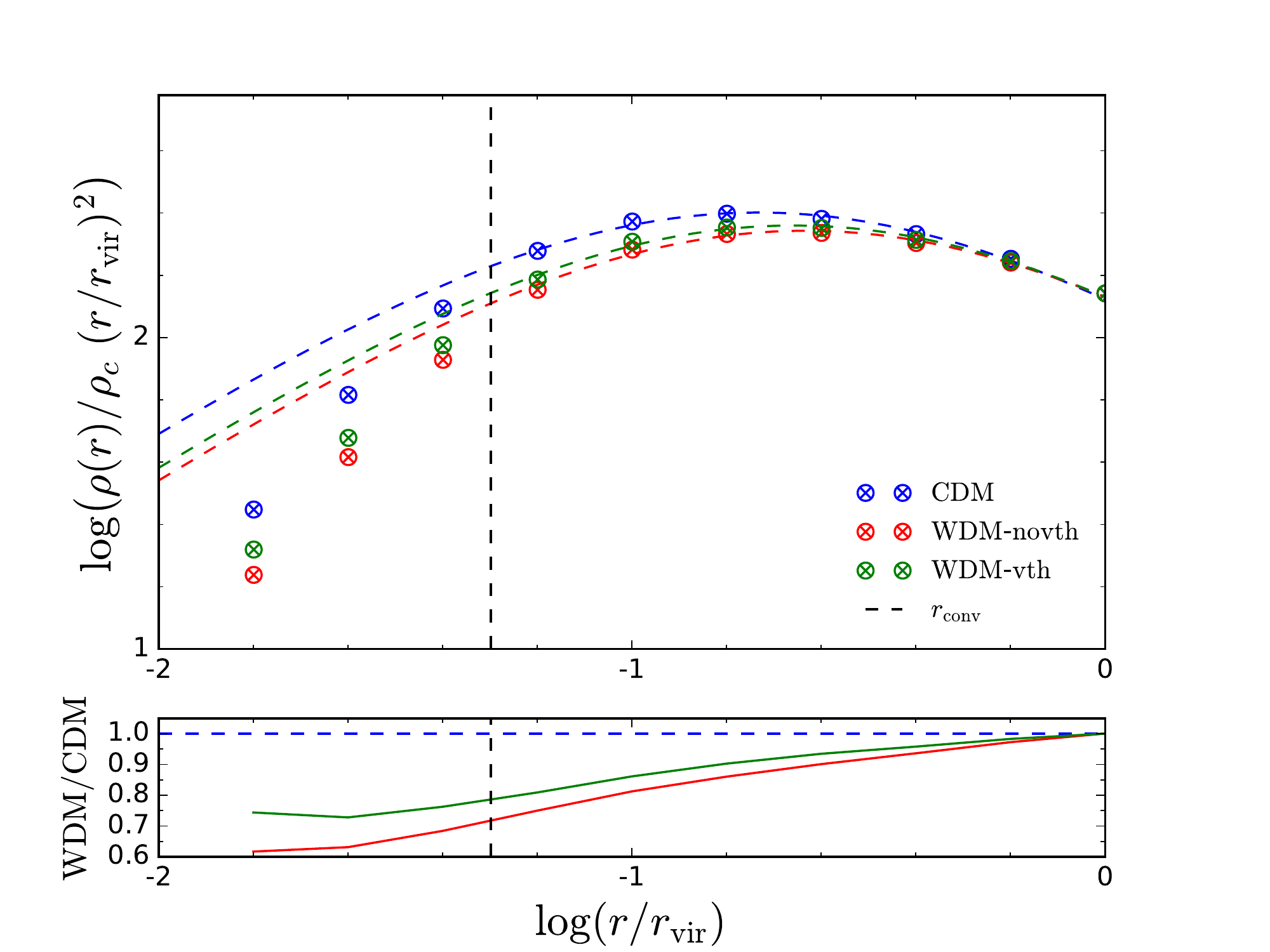}\label{fig:1c2haloesok}} \\   

\caption{Some halo radial density profiles extracted from CDM (blue), $\mathrm{WDM}${-}$\mathrm{vth}$ (green) and $\mathrm{WDM}${-}$\mathrm{novth}$ (red) simulations at $z=0$. (a-b) WDM with mass $m_\mathrm{WDM} =3.3$ keV. (c-d) WDM with mass $m_\mathrm{WDM} =2$ keV. The coloured dashed curves represent the NFW fits of the radial density profiles measured from simulations. The black dashed line displays the position of the convergence radius $r_\mathrm{conv}$. The bottom panel of each figure shows the ratio between the WDM and CDM radial profiles measured from our simulations.}
\label{fig:33haloesok}
\end{figure}

In \cite{Bose:2015mga} it was found that WDM radial density profiles also follow a NFW form in the outer parts of the halo. We examine this feature in figure \ref{fig:33haloesok}, where we display some examples of radial density profiles measured from simulations at $z=0$, stacking haloes in mass bins centred on $\log (M^\mathrm{central}_\mathrm{vir}/h^{-1}\text{M}_{\odot}) = 10, 11.8$ with width $0.2$ dex. We plot the median density for the haloes in each radial bin. The radial bins are chosen such that the logarithmic difference between the central radius values ($r_i$) for two near bins is $|\log(r_i/r_\mathrm{vir}) - \log(r_{i+1}/r_\mathrm{vir})|=0.2$ with $i\in\{1,N_\mathrm{bin}\}$ and $N_\mathrm{bin}$ is the total number of bins. We fit the NFW formula using each of these dark matter profiles between the convergence radius\footnote{The convergence radius is defined to be the radius within which the relaxation time is of the order of the age of the Universe \cite{Power:2002sw} and it is intended as the radius beyond which the halo mass density profile is reliably modelled for a given halo mass for a given simulation set-up.} $r_\mathrm{conv}$ and the virial radius $r_\mathrm{vir}$. The best fit is found by minimising the following sum of squares:
\begin{equation}
Q_{N_\mathrm{bin}}(\delta_c,r_s) = \frac{1}{N_\mathrm{bin}}\sum^{N_\mathrm{bin}}_{i=1} \left[ \log\left(\frac{\rho_i}{\rho_c}\left(\frac{r_i}{r_\mathrm{vir}}\right)^2  \right)  - \log\left(\frac{\rho_{\mathrm{fit}}(r_i,\delta_c,r_s)}{\rho_c}\left(\frac{r_i}{r_\mathrm{vir}}\right)^2  \right) \right]^2,
\end{equation}
where $\rho_{i}$ is the value of the radial density profile extracted from our simulations at the radius $r_i$, while $\rho_{\mathrm{fit}}(r_i,\delta_c,r_s)$ is the value of the NFW fit in eq. (\ref{eq:NFW}) at the same radius $r_i$.
As we can see from figure \ref{fig:33haloesok}, the halo radial density profiles extracted from our simulations of CDM and WDM both agree quite well with the NFW fits. Moreover, in the larger halo mass bin, the density profiles in the WDM simulations are not significantly affected by the small scale cut-off in the initial power spectrum, so they do not present significant differences with respect to those measured from CDM. This is true for both the WDM candidates with $m_\mathrm{WDM}=3.3$ and 2 keV, as can be seen from figures \ref{fig:1a33haloesok} and \ref{fig:1a2haloesok} respectively. The effect of adding thermal velocities to the initial conditions is also very small for high mass haloes. In the smaller halo mass bin, on the other hand, the density profiles in WDM simulations are clearly shallower than in CDM towards the halo centres -- $\sim10\%$ for the case of $3.3$ keV WDM particles and $\sim30\%$ for the $2$ keV case near the convergence radius. The effect is more prominent in $\mathrm{WDM}${-}$\mathrm{novth}$ than in $\mathrm{WDM}${-}$\mathrm{vth}$. This means that the numerical noise caused by including thermal velocities modifies the slope of the radial density profiles for the small halo masses probed in our analysis.

\begin{figure}
\centering
\subfigure[][$m_\mathrm{WDM} =  2$ keV, $L = 25$ $h^{-1}$Mpc]
   {\includegraphics[width=.52\textwidth]{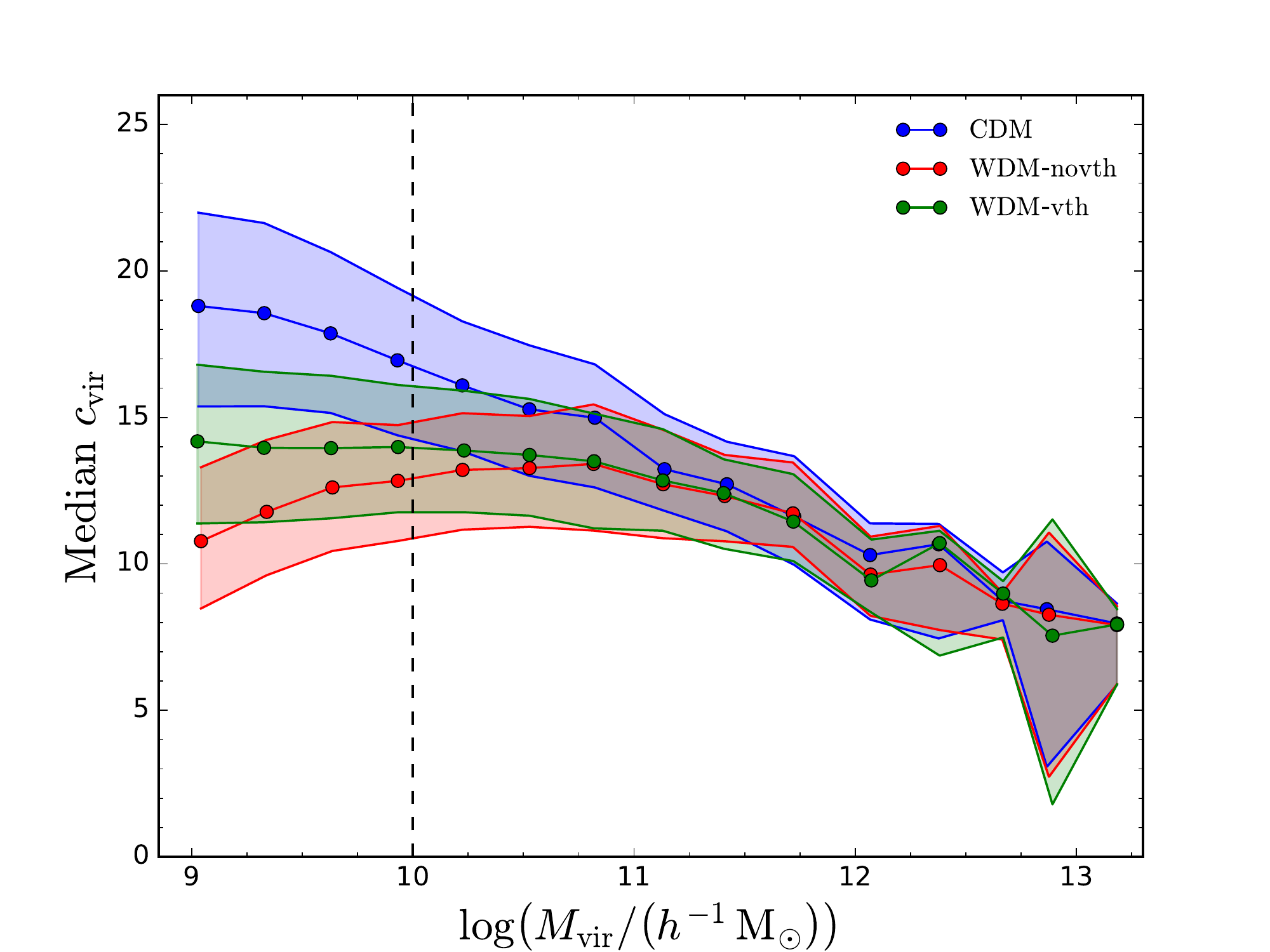}\label{fig:1b2keV}} \hspace{-2.2\baselineskip} 
\subfigure[][$m_\mathrm{WDM} =  3.3$ keV, $L = 25$ $h^{-1}$Mpc]
   {\includegraphics[width=.52\textwidth]{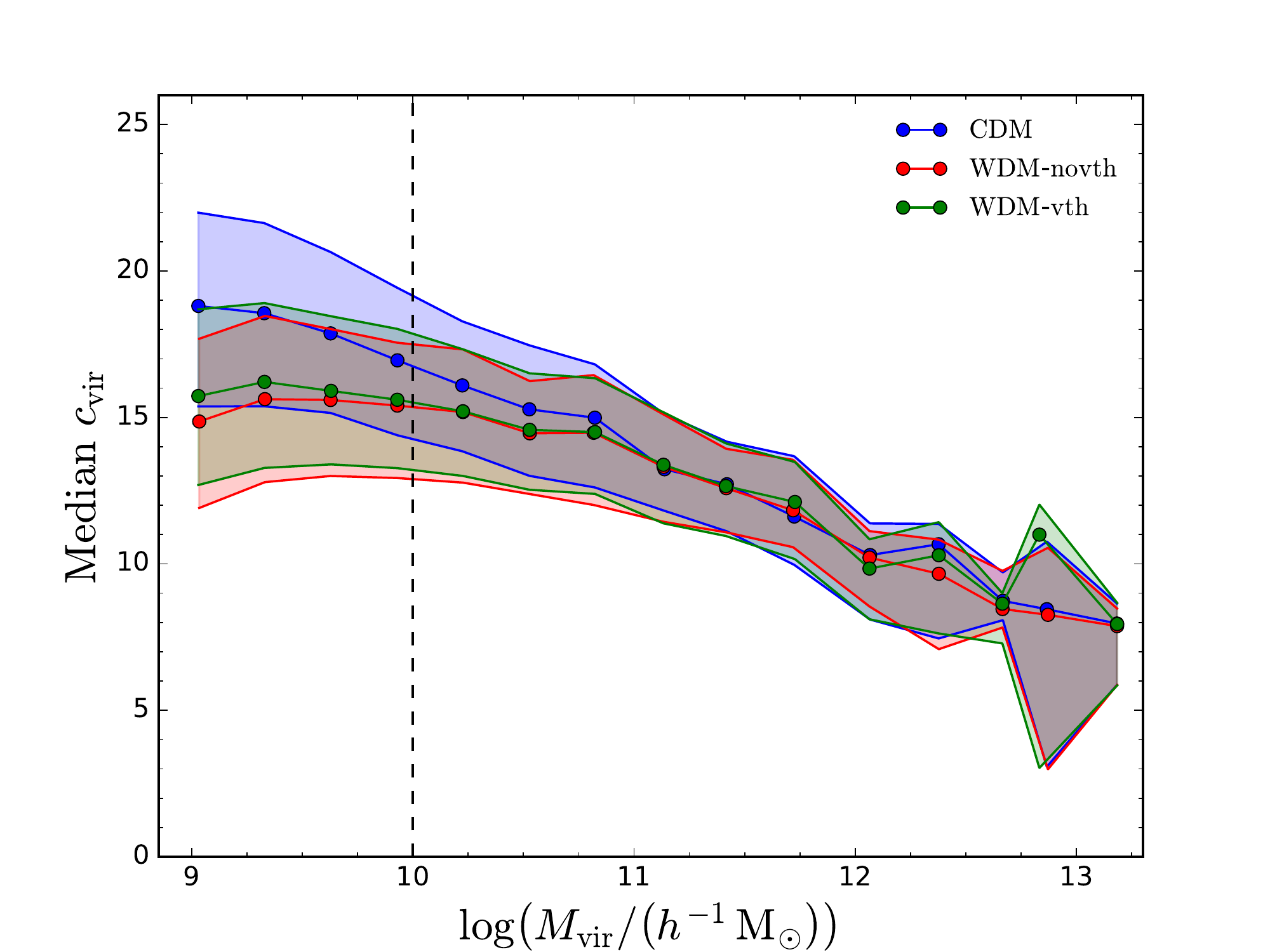}\label{fig:1b33keV}} \\
\subfigure[][$m_\mathrm{WDM} =  7$ keV, $L = 25$ $h^{-1}$Mpc]
   {\includegraphics[width=.52\textwidth]{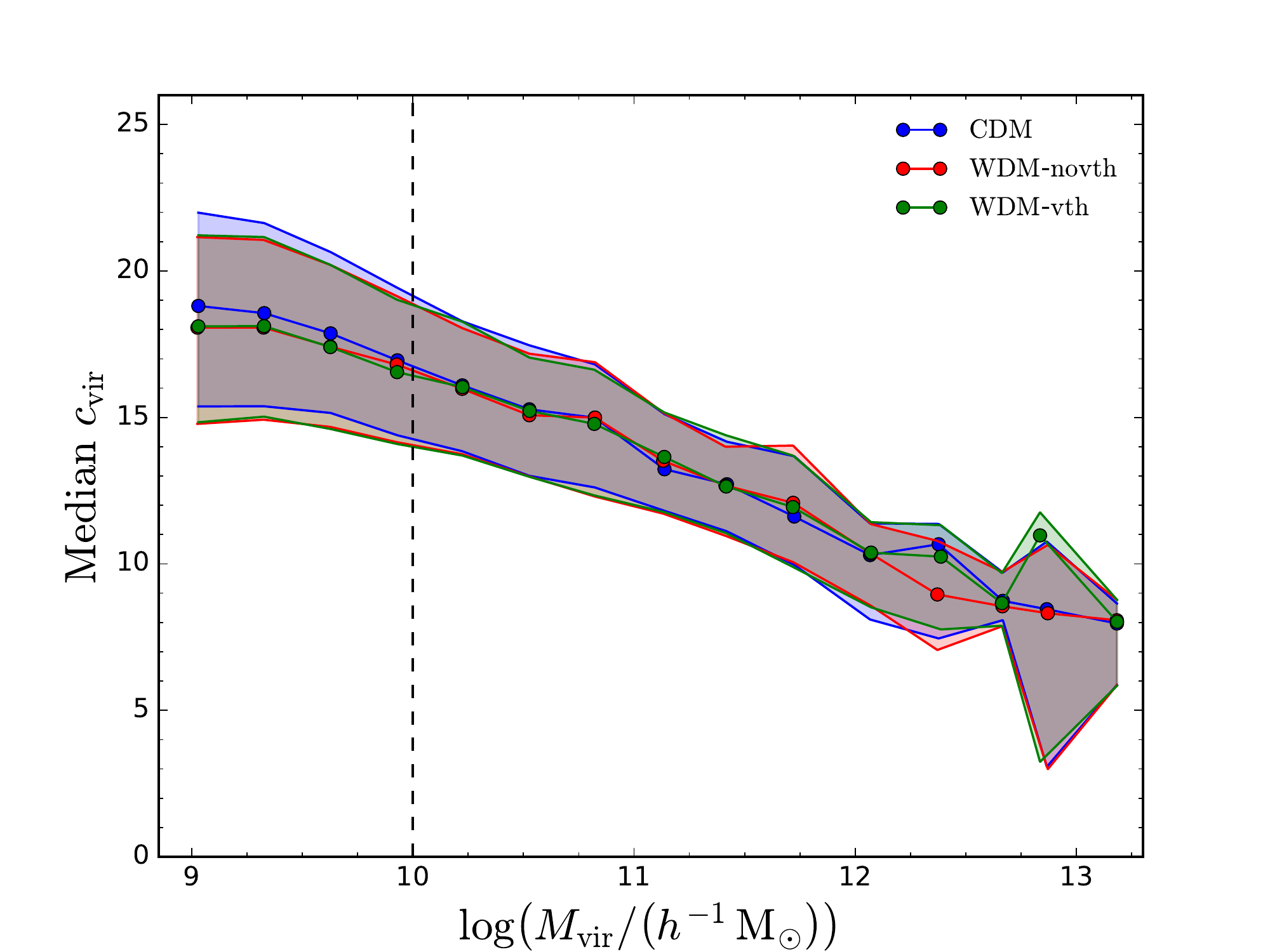}\label{fig:1bl25}} \hspace{-2.2\baselineskip}
\subfigure[][$m_\mathrm{WDM} =  7$ keV, $L = 12$ $h^{-1}$Mpc]
   {\includegraphics[width=.52\textwidth]{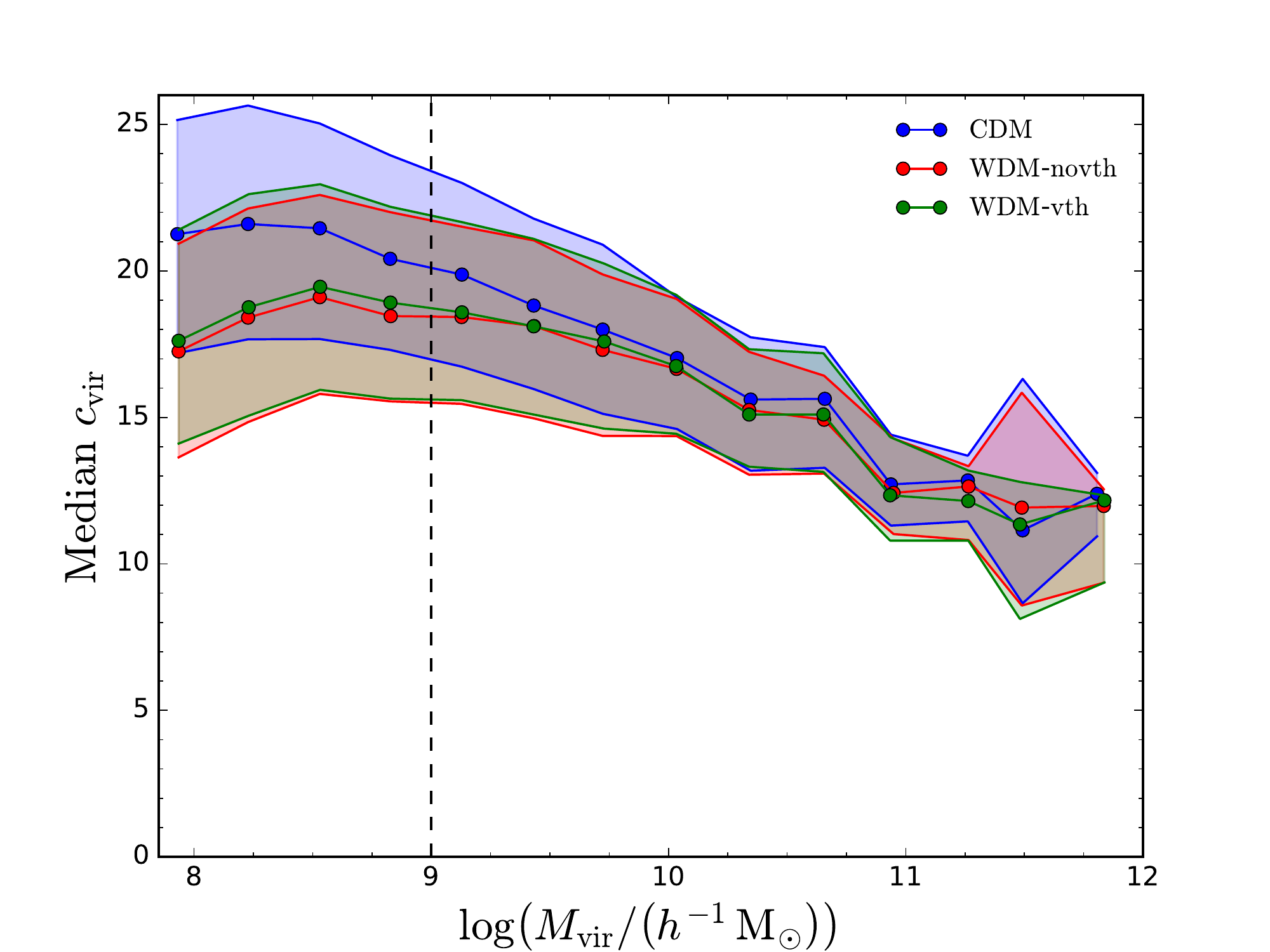}\label{fig:1bl12}} \\   
\caption{Median concentration parameter $c_\mathrm{vir}$ for CDM and WDM with $m_\mathrm{WDM} =  2,3.3,7$ keV at redshift $z=0$. For the WDM candidate mass $m_\mathrm{WDM}= 7 $ keV we present also the results from the high resolution simulation with  $L = 12$ $h^{-1}$Mpc. The bands show the $68\%$-ile range around the median. The black vertical dashed line indicates haloes with mass $\sim1000$ times the mass of our simulation particles, which is $M_\mathrm{sim} = 1.01 \times 10^7 h^{-1}\mathrm{M}_{\odot}$ for $L = 25$ $h^{-1}$Mpc and $M_\mathrm{sim} = 1.11 \times 10^6 h^{-1}\mathrm{M}_{\odot}$ for $L = 12$ $h^{-1}$Mpc.}
\label{fig:concentration}
\end{figure}

The concentration parameter, $c_\mathrm{vir} \equiv r_\mathrm{vir}/r_{s}$, is usually used to quantify the steepness of the inner density profiles of NFW haloes. Figure \ref{fig:concentration} shows the values of this parameter at redshift $z=0$ for our CDM and WDM simulations. At large halo masses, the concentrations measured from WDM and CDM simulations agree well between each other. However, there are differences at lower masses. While the concentration parameter continues to increase in CDM, it turns over and decreases in $\mathrm{WDM}${-}$\mathrm{novth}$ simulations at small masses (this is well known in the literature, see e.g. \cite{Bose:2015mga,2012MNRAS.424..684S}). The mass scale at which the downturn appears depends on the WDM candidate mass. For example, the downturn appears at masses $\lesssim 10^{11}$ $h^{-1}$M$_\odot$ for the WDM candidate with mass $m_\mathrm{WDM}=2$ keV (see figure \ref{fig:1b2keV}), while it is shifted to smaller masses, $\lesssim 3\times10^{8}$ $h^{-1}$M$_\odot$, for the case of $m_\mathrm{WDM}=7$ keV (see figure \ref{fig:1bl12}).

In $\mathrm{WDM}${-}$\mathrm{vth}$ the reduction in the concentration is less pronounced than in $\mathrm{WDM}${-}$\mathrm{novth}$; indeed the median values extracted from $\mathrm{WDM}${-}$\mathrm{vth}$ are in general larger than those measured from $\mathrm{WDM}${-}$\mathrm{novth}$ at small halo masses.  As expected the magnitude of these differences depends on the mass of the WDM candidate. For instance, for masses around $m_\mathrm{WDM} =2$ keV, $c_\mathrm{vir} \simeq 13.9 \pm 2.2$ (the $68\%$-ile range around the median) in $\mathrm{WDM}${-}$\mathrm{vth}$  for haloes with masses of $10^{10}$ $h^{-1}$M$_\odot$, while the value predicted from $\mathrm{WDM}${-}$\mathrm{novth}$ is $c_\mathrm{vir} \simeq 12.8$. For colder WDM candidates the effect is weaker at the mass scales probed by our simulations; indeed for the case of particle mass $m_\mathrm{WDM}=7$ keV the effect is negligible for simulations with $L = 25$ $h^{-1}$Mpc (see figure \ref{fig:1bl25}), while some differences can be noticed at the very small masses in simulation with $L = 12$ $h^{-1}$Mpc (see figure \ref{fig:1bl12}), however the values of $c_\mathrm{vir}$ measured from $\mathrm{WDM}${-}$\mathrm{novth}$ and $\mathrm{WDM}${-}$\mathrm{vth}$ simulations are always compatible between each other within the $68\%$-ile range around the median.

\begin{figure}
\centering
\subfigure[][ICs at $z_\mathrm{ini}=199$.]
   {\includegraphics[width=.52\textwidth]{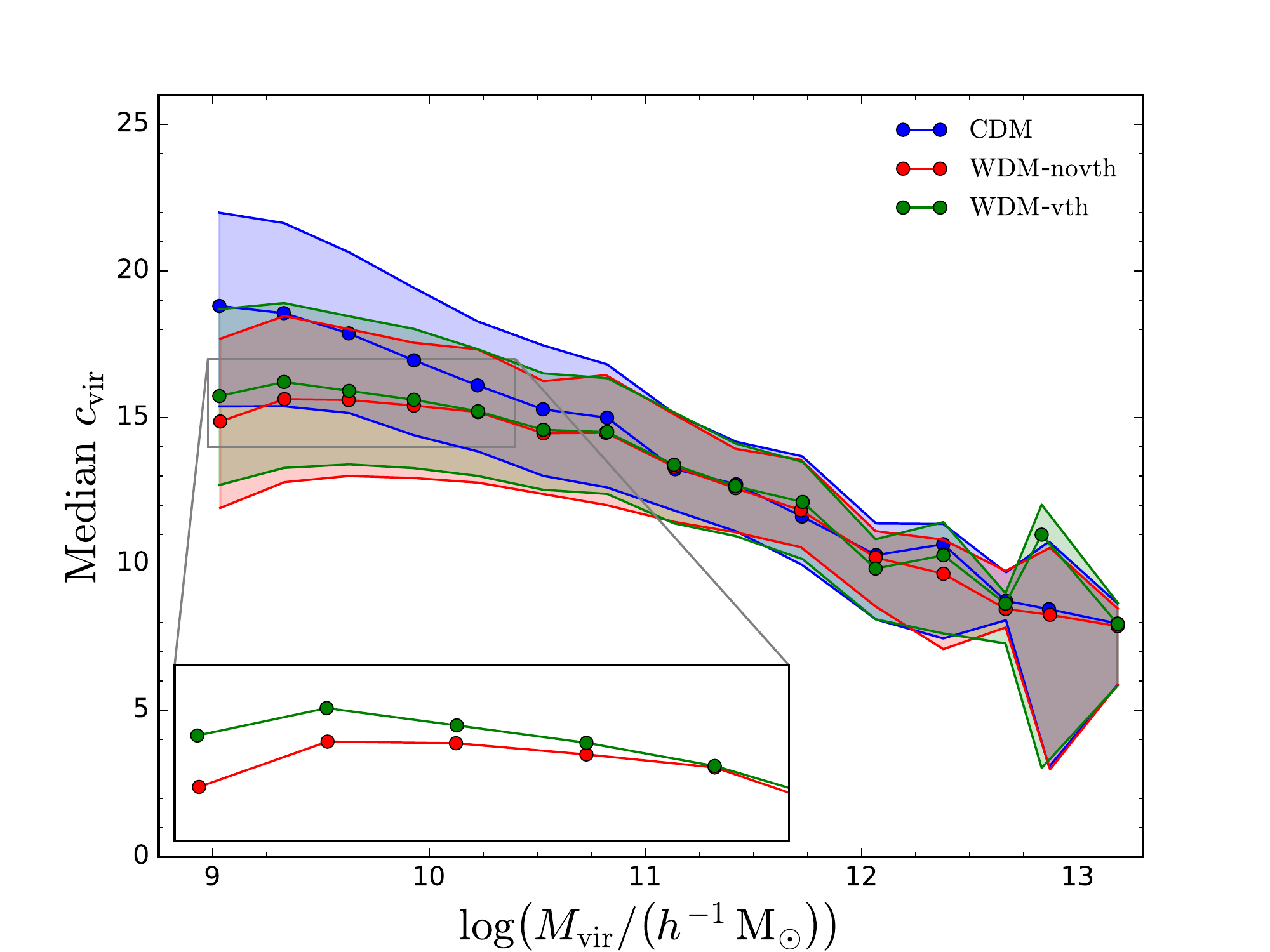}\label{fig:1b}} \hspace{-2.\baselineskip}
\subfigure[][ICs at $z_\mathrm{ini}=39$.]
   {\includegraphics[width=.52\textwidth]{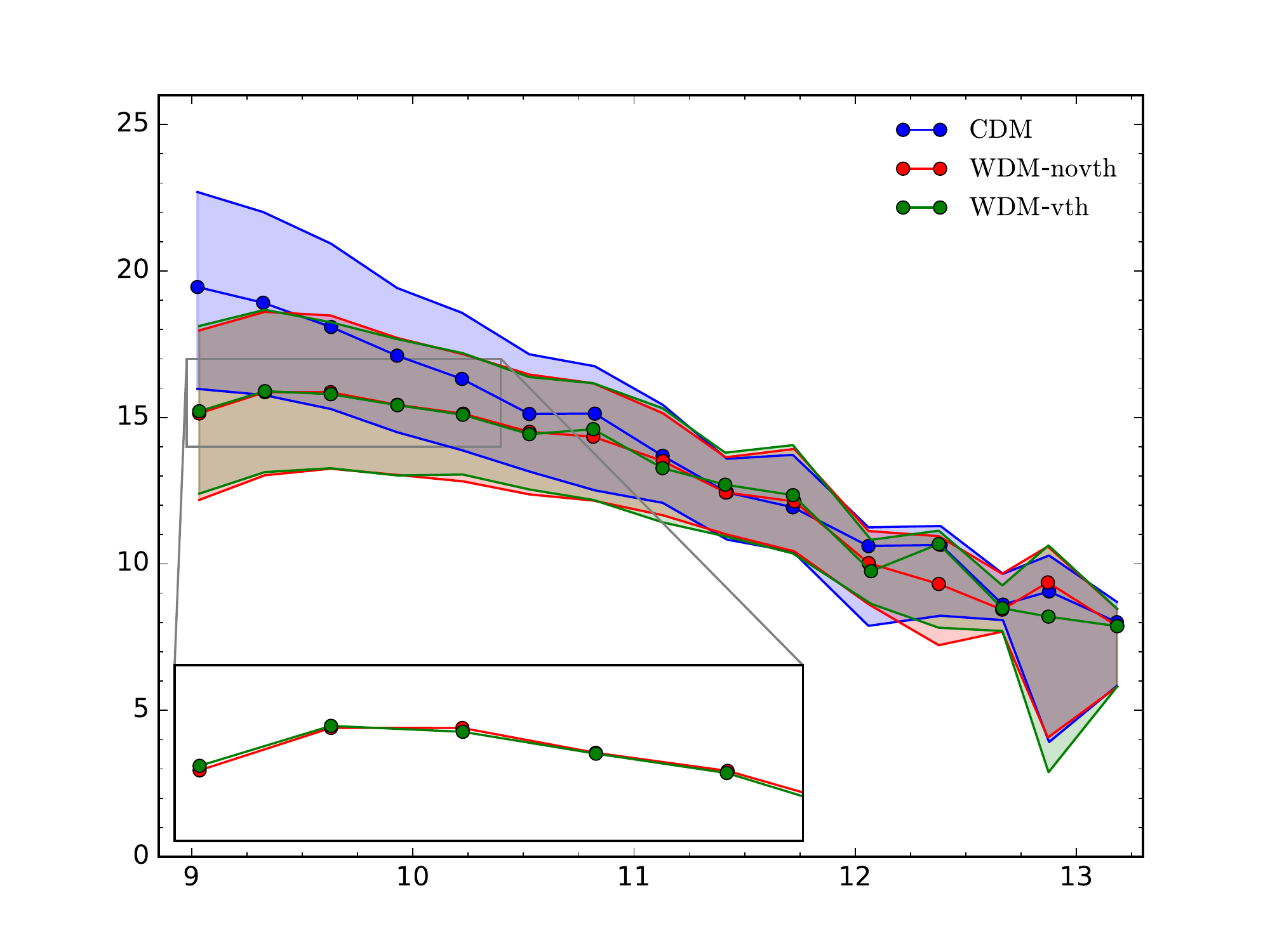}\label{fig:1c}} \quad   
\caption{Median concentration parameter $c_\mathrm{vir}$ at $z=0$ for CDM and WDM with $m_\mathrm{WDM} =  3.3$ keV for simulations with (a) ICs at $z_\mathrm{ini}=199$ and (b) ICs at $z_\mathrm{ini}=39$. The bands show the $68\%$-ile range around the median.}
\label{fig:concentrationz39}
\end{figure}

For completeness, we have compared the halo concentration-mass relations for WDM and CDM simulations starting at different initial redshifts ($z_\mathrm{ini}=199$ on the left and $39$ on the right), for the case of $m_\mathrm{WDM}=3.3$ keV, in figure \ref{fig:concentrationz39}. The differences between $\mathrm{WDM}${-}$\mathrm{vth}$ and $\mathrm{WDM}${-}$\mathrm{novth}$ seen in simulations with initial redshift $z_\mathrm{ini}=199$ are not present in those starting at $z_\mathrm{ini}=39$. This confirms that the observed deviations in the halo properties of simulations with thermal velocities are due to numerical artefacts. 

As studied in \cite{Hogan:2000bv,2013MNRAS.428..882M,2013MNRAS.430.2346S}, imprinting primordial thermal velocities on the particles ensures a ``phase-packing'' limit, which prevents the density in the central region of the haloes from increasing arbitrarily, producing a central core.  However, for values of WDM candidate masses compatible with the upper limits from the Ly-$\alpha$ forest \cite{Viel:2013apy, Irsic:2017ixq}, the cores are only a few parsecs in size and not astrophysically relevant \cite{2013MNRAS.428..882M,2013MNRAS.430.2346S}. This explains why we do not expect to find such new physical features in halo properties when adding thermal velocities.

\section{Conclusions}

Several non-cold DM scenarios have been proposed to solve the small-scale problems of the standard cosmological paradigm. Among them, the simple model of thermal WDM is the most studied \cite{Bode:2000gq,Colin:2000dn,Hansen:2001zv,Viel:2005qj,Viel:2013apy,Irsic:2017ixq,Boyarsky:2008xj,Schneider:2013ria,2013MNRAS.428..882M,2013MNRAS.430.2346S,Paduroiu:2015jfa,Colin:2007bk,Wang:2007he,Lovell:2013ola,2012MNRAS.424..684S,Bose:2015mga,Power:2013rpw,Power:2016usj}. Thermal WDM candidates are characterised by Fermi-Dirac distributions for the thermal velocities.  However, apart from the induced free-streaming effect, the role of thermal velocities in structure formation remains unclear. We have investigated the effect of thermal velocities in N-body simulations of structure formation in WDM cosmologies when the thermal velocities are imposed on the simulation particles. 

At high redshift, $z=199$, thermal velocity dispersions are non-negligible with respect to the peculiar velocities and they must be taken into account in the initial conditions. However, when this is done, a new source of numerical noise affects the simulations. We have improved upon the results presented in \cite{Colin:2007bk} by increasing the simulation resolution to reach the scales relevant to free-streaming. The results in \cite{Colin:2007bk} are based on a WDM candidate (a non-thermal sterile neutrino with mass $0.5$ keV), which is now ruled out by the current Ly-$\alpha$ constraints \cite{Viel:2013apy, Irsic:2017ixq}. Here, we have focused on more realistic WDM masses, $m_\mathrm{WDM} \geq 2$ keV, which are compatible with the limits from the Ly-$\alpha$ forest.  Moreover, we have extended the analysis of \cite{Colin:2007bk} by measuring the velocity power spectra, spanning a vast range of redshifts, from early times ($z=199$) to the present day and we have estimated the impact of the noise at different times. The initial velocity power spectrum in simulations with thermal velocities is always affected by the noise caused by adding thermal velocities to the peculiar velocities of simulation particles. The noise propagates to all subsequent times, influencing the matter and velocity power spectra. These numerical artefacts dramatically affect the simulation predictions at early times, when all the modes are in the linear regime, producing enhancements in the velocity power of $\sim 100$ (for WDM candidates with masses of $3.3$ keV) and of $\sim 3000$ (for  masses of $2$ keV) at the Nyquist frequency $k_{Ny} \sim 64$ $h/$Mpc. At late times, these effects are less pronounced because the non-linear growth of structure gradually dominates over the noise. Indeed, at $z\leq2$ the deviations do not exceed $6\%$ in the velocity spectrum and $2\%$ in the matter spectrum for scales $10 <k< 64$ $h/$Mpc for masses of $3.3$ keV. However, for the masses around $2$ keV there are still appreciable deviations in the velocity power spectrum at $z\leq2$ around $18\%$.

Looking at halo properties at late times, such as the halo mass function and the radial density profile of haloes, we found some differences in simulations with thermal velocities with respect to those without. These deviations are due to the noisy initial conditions of the simulations which include thermal velocities. Due to this new source of noise, more spurious substructures appear and the cut-off mass proposed by \cite{Wang:2007he} fails to reproduce the mass at which the upturn in the halo mass function appears due to spurious haloes. We find a new mass cut-off, $M_{\text{lim}} = 32.2\, \bar{\rho}\, d\, k_{\text{peak}}^{-2}$, which works better at reproducing the mass scale below which spurious haloes start to be important in simulations with thermal velocities.  

The standard practice of imposing thermal velocities in N-body simulations by adding them to the peculiar velocities in the initial conditions then reduces the range of validity of simulation predictions with respect to simulations without thermal velocities because of the noise introduced in the initial conditions. The noise can be reduced by increasing the number of simulation particles or starting at a lower redshift. Obviously, both of the options introduce side effects: increasing the number of particles increases the computational time needed to evolve the simulations, while by starting at very low redshifts, we inevitably lose information about the high redshift behaviour of the system and the accuracy of the calculation could also be reduced. Our results help to determine the range of applicability of simulations of WDM.

We have focused on the simple scenario of thermal WDM, where warm particles account for all the DM content. These models are characterised by strong cut-offs and WDM candidate masses around keV, both of these properties prevent significant physical effects of thermal velocities. Indeed, thermal velocities only affect the inner-most parts of the haloes \cite{2013MNRAS.428..882M,2013MNRAS.430.2346S}. The situation changes if we consider composite DM models (e.g. mixed DM or CDM + massive $\nu$s), such models have no strong cut-offs because of the remaining CDM influence once the WDM component has died off \cite{Boyarsky:2008xj}. This means that we can have haloes down to very small scales, where the thermal motion can be important \cite{Brandbyge:2008rv, Viel:2010bn,2013JCAP...03..019V}. Composite DM simulations can then clarify the physical role of thermal velocities in structure formation, for this reason they deserve further investigation. The analysis conducted in the present work can be then considered as a guide when approaching the problem of adding thermal velocities in N-body simulations for a generic DM model: if they are included as we did in our work, the simulations need to be treated with extra care and a resolution analysis is necessary to distinguish physical effects (if they are present) from numerical noise.
\section*{Acknowledgments}

We thank Sownak Bose, Marius Cautun, Oliver Hahn, Jianhua He, Chris Power and Matteo Viel for valuable discussions. ML and SP are supported by the European Research Council under ERC Grant ``NuMass'' (FP7- IDEAS-ERC ERC-CG 617143). BL is supported by an European Research Council Starting Grant (ERC-StG-716532-PUNCA). CMB and BL acknowledge the support by the UK STFC Consolidated Grants (ST/P000541/1 and ST/L00075X/1) and Durham University. This work used the DiRAC Data Centric system at Durham University, operated by the Institute for Computational Cosmology on behalf of the STFC DiRAC HPC Facility (\href{www.dirac.ac.uk}{www.dirac.ac.uk}). This equipment was funded by BIS National E-infrastructure capital grant ST/K00042X/1, STFC capital grants ST/H008519/1 and ST/K00087X/1, STFC DiRAC Operations grant  ST/K003267/1 and Durham University. DiRAC is part of the National E-Infrastructure.

\end{document}